\documentclass[iop]{emulateapj}

\shorttitle{Three dimensional maps of the LMC}
\shortauthors{Haschke, Grebel \& Duffau}
\slugcomment{submitted: 20 July 2011; accepted: 13 July 2012}

\begin{document}
 \title{Three dimensional maps of the Magellanic Clouds \\using RR~Lyrae stars and Cepheids \\I. The Large Magellanic Cloud}


\author{Raoul Haschke\altaffilmark{*}
 ,
	 Eva K. Grebel,
	 and
 Sonia Duffau
	 }

\email{haschke@ari.uni-heidelberg.de}

\affil{Astronomisches Rechen-Institut, Zentrum f\"ur Astronomie der Universit\"at Heidelberg,
	 M\"onchhofstrasse 12--14, D-69120 Heidelberg, Germany}\

\altaffiltext{*}{Raoul Haschke is a member of the Heidelberg Graduate School for Fundamental Physics (HGSFP) and of the International Max Planck Research School for Astronomy and Cosmic Physics at the University of Heidelberg.}

\begin{abstract}
The new data for Cepheids and RR~Lyrae stars of the Optical Gravitational Lensing Experiment (OGLE-III) survey allow us to study the three-dimensional distribution of stars corresponding to young (a few tens to a few hundreds of millions of years) and old (typically older than $\sim$ 9~Gyr) populations of the Large Magellanic Cloud
(LMC) traced by these variable stars. 
We estimate the distance to 16949 RR~Lyrae stars by using their photometrically estimated metallicities. Furthermore the periods of 1849 Cepheids are used to determine their distances. Three-dimensional maps are obtained by using individual reddening estimates derived from the intrinsic color of these stars. The resulting median distances of the RR~Lyrae stars and Cepheids appear to resolve the long and short distance scale problem for our sample. With median distances of $53.1 \pm 3.2$~kpc
for the RR~Lyrae stars and $53.9 \pm 1.8$~kpc for the Cepheids, these two distance indicators are in very good agreement with each other in contrast to a number of earlier studies. Individual reddening estimates allow us to resolve the distance discrepancies often observed while comparing Cepheids and RR~Lyrae stars. For both
stellar populations we find the inclination angle of the LMC to be $32 \pm 4^\circ$ and the mean position angle to be $115 \pm 15^\circ$. The position angle increases with galactocentric radius, indicative of mild twisting. Within the innermost 7~degrees of the LMC covered by OGLE~III the change in position angle amounts to more than 10~degrees. The depth of the Cepheids is found to be $1.7 \pm 0.2$~kpc. The bar stands out as an overdensity both in RR~Lyrae stars and in Cepheids. In RR~Lyrae stars the bar can be traced as a protruding overdensity with a line-of-sight depth of almost 5~kpc in front of the main body of the disk.

\end{abstract}


%
\defcitealias{Haschke11_reddening}{Paper~I}
\defcitealias{Haschke12_MDF}{Paper~II}
\defcitealias{Zinn84}{ZW84} 
\defcitealias{Jurcsik95}{J95} 
%

%

\section{Introduction}
\label{introduction}

The Large Magellanic Cloud (LMC) is the largest and one of the closest satellites of the Galaxy. This irregular galaxy has long been a focus of astronomical studies. Because of its proximity it is a popular object to study star and cluster formation and evolution at lower metallicities than in the Milky Way \citep[e.g.,][, to just name a few]{Walborn99, Glatt10, Cioni09, Johnson06}. The LMC is part of an interacting galaxy triplet along with the Small Magellanic Cloud (SMC) and with the Milky Way; an interaction that may have a substantial effect also on the Milky Way \citep[e.g.,][]{Weinberg06}. Interestingly, such Magellanic-Cloud-like satellites are rarely found around massive galaxies \citep{James11, Liu11}. However, it is unclear for how long the Magellanic Clouds and the Milky Way have been together and whether the LMC is actually a bound satellite \citep[e.g.,][]{Besla07, Bekki09}. 

Moreover, the LMC is a prime target for studies attempting to calibrate the cosmological distance scale, in particular with respect to (classical) Cepheid distances \citep[e.g.,][]{Freedman01}. Distances to both Clouds have been determined with many different techniques based on, e.g., variable sources such as eclipsing binaries, supernova 1987A, novae, Type 2 Cepheids, Miras, $\delta$ Scuti stars, RR Lyrae of type {\em ab} or of type {\em c}), non-variable groups of specific types of stars (e.g., red clump stars, early-type stars, tip-of-the-red-giant-branch stars), or isochrone fits to entire stellar populations such as star clusters or field main-sequence stars. Summaries of the methods and their various results are given in, e.g., \citet{Westerlund97}, \citet{Alves04}, and \citet{Schaefer08}. 

As noted in these studies, past determinations ranged from a ``short'' LMC distance scale with a distance modulus around 18.1 to 18.2 mag to a ``long'' distance scale with a distance modulus around 18.7 to 18.8. Regarding two of the most widely distance indicators, RR~Lyrae stars often resulted in closer LMC distances, whereas Cepheid measurements often placed the Clouds at somewhat larger distances. Most of the recent studies though yield distance moduli between 18.45 and 18.65 \citep[see][for a critique of this apparenly good agreement]{Schaefer08}. As noted by \citet[][and references therein]{McNamara11}, the seemingly larger Cepheid distances may be caused by Cepheids being located in more highly extincted areas because of their younger age than the other variables mentioned above. Other factors may include uncorrected effects of LMC geometry, a metallicity dependence of the period-luminosity relation of Cepheids, or truly different distances of the different distance indicators \citep[e.g.,][]{Westerlund97, Feast99}.

With an estimated age range of $\sim 30$ -- $\sim 300$~Myr \citep{Grebel98}, Cepheids are excellent tracers of the young stellar population. RR Lyrae stars have ages older than 9 Gyr and thus are good tracers of the old population in galaxies. Owing to their well-defined period-luminosity-metallicity relation they are valuable standard candles if their metallicity is known. For a more detailed overview of the distance estimates using RR Lyrae stars and Cepheids as well as other tracers, we refer to the book by \citet{Westerlund97} and the articles by \citet{Feast99, Benedict02, Alves04, Matsunaga11}, and \citet{Walker11}. 
 
A variety of methods and objects have been used to investigate the structure of the Magellanic Clouds (MCs). At optical and infrared wavelengths large surveys reveal that the stellar density distributions and locations of younger and older stellar populations are quite different \citep[e.g.,][] {Cioni00a, Marel01a, Weinberg01, Zaritsky04, Lah05, Glatt10, Subramanian10}, and their centroids differ from those of the neutral hydrogen \citep{Staveley03}. The orientation and three-dimensional shape of the LMC have been studied repeatedly. For red giants \citet{Marel01b} found a position angle $\theta = 122.5^\circ \pm 8.3^\circ$ in good agreement with \citet{Subramaniam09a}, who analyzed RR~Lyrae stars from the Optical Gravitational Lensing Experiment (OGLE-III) \citep{Soszynski09}. Using the same data, \citet{Pejcha09}
published a position angle consistent with the value published by \citet{Subramaniam04} of $\theta = 114^\circ \pm 22^\circ$ for red clump (RC) stars. Similarly, using data from the VISTA near-infrared survey of the Magellanic system (VMC), \citet{Rubele12} obtained $\theta = 129.1^\circ \pm 13.0^\circ$. In contrast to these determinations \citet{Nikolaev04} found $\theta = 154^\circ \pm 3^\circ$ with Cepheids from the MACHO survey. This is in agreement with the work of \citet{Koerwer09} using the RC of the LMC. 

The distance measurements of the LMC suggest that the eastern part is closer to the Sun than the western part. The inferred inclination angle varies between $i = 23.0^\circ \pm 0.8^\circ$ and $i = 37.4^\circ \pm 2.3^\circ$ \citep{Subramanian10}. For instance, the studies by \citep{Marel01a}, \citet{Lah05}, \citet{Nikolaev04},
\citet{Koerwer09}, \citet{Subramaniam09a}, and \citet {Rubele12} derive inclination angles within this range using various types of stellar tracers. \citet{Subramanian10} find evidence for a warped disk. Using velocities of red supergiants, AGB stars, and giants, \citet{Olsen11} suggest that stars are currently being accreted from the SMC and that these stars form a kinematically distinct population.

The OGLE survey looks for microlensing events in the Galactic center and in the MCs. Therefore a considerable portion of the dense regions of the MCs was monitored for many years \citep{Udalski92, Udalski97, Udalski08a}. The new OGLE~III data release combines six years of observations with the largest field coverage in the MCs ($\sim$
40~degrees) of the OGLE experiment obtained thus far. 

RC stars observed in OGLE~III indicate a line-of-sight depth of 4~kpc for the LMC bar and 3.4~kpc for the disk \citep{Subramanian09}, in good agreement with \citet{Clementini03} evaluating RR~Lyrae stars. 

\citet{Zhao00} discussed arguments for a bar located in front of the LMC. These results were confirmed by \citet{Nikolaev04} using Cepheids in the LMC. They showed that the bar is at least 0.5~kpc in front of the disk. Model fitting based on the data of the Magellanic Cloud Photometric Survey by \citet{Zaritsky04b} obtained a similar result as \citet{Nikolaev04}. In contrast to these results \citet{Subramaniam09b} did not find evidence for a bar in front of the LMC investigating RC stars from the OGLE~III data. \citet{Bekki09b} proposed the collision of a dark halo with the LMC as the origin for a bar located in front of the main body of the LMC. \citet{Zhao00} speculate that the bar could either be ``a tidally stretched companion of the LMC, originating from the proto-Magellanic Cloud'' or a dynamically young feature possibly induced by tidal interactions.

We use the entire sample of OGLE RR~Lyrae stars and Cepheids, covering a field much larger than in previous studies of these objects, to obtain three-dimensional maps of these old and young populations of the LMC. The distances are derived taking into account star-by-star extinction corrections. Such maps for different populations, in combination with kinematics, will ultimately also help us to understand the evolution of the LMC better. 

In Section~\ref{data} we discuss the data released by the OGLE collaboration and used in our study. Distance estimations and reddening corrections are discussed in Section~\ref{distance}. We then first analyze the spatial distribution of our stars projected in two dimensions in Section~\ref{Density_of_OGLEIII}, before using the metallicities of the individual RR~Lyrae stars determined in \citet[from now on Paper~II]{Haschke12_MDF}, in order to calculate the reddening-corrected distances in Section~\ref{3D}. In Section~\ref{3D_structure} structural characteristics such as the inclination angle, position angle, and scale height of the LMC are presented. We summarize and discuss our results in Section~\ref{Conclusions}.

%

\section{Data}
\label{data}

Our paper is based on public data from the OGLE time-domain imaging survey. During its phase~I, the OGLE experiment collected imaging data from 1992 to 1995 with the 1~m Swope telescope of Las Campanas Observatory in Chile \citep{Udalski92}. Since 1995, with the start of phase~II \citep{Udalski97}, OGLE's own 1.3~m telescope at Las Campanas became operational. During its first two phases, the OGLE collaboration used a $2048 \times 2048$~pixel camera with a field of view of $15' \times 15'$. Imaging data were taken in the $B$, $V$, and, preferentially, the $I$ band and covered 4.5 square degrees of the central parts of the LMC. 

In 2001 the third phase (OGLE~III) began. OGLE took data until mid-2009 using a mosaic camera consisting of eight CCDs with $2048 \times 4096$ pixels each. With a total detector size of $8192 \times 8192$~pixels the OGLE~III field of view covers an area of $35' \times 35'$ at once \citep{Udalski03}. Altogether OGLE~III covers nearly 40 square degrees of the LMC disk and bar. These observations provide photometry in the $V$ and $I$ bands for about 35~million stars \citep{Udalski08a}. Apart from the full photometric catalog the OGLE collaboration provides subsets of their data products for, e.g., RR~Lyrae stars or high proper motion stars, in separate catalogs. Each catalog of the LMC \footnote{The catalogs are available from \url{http://ogle.astrouw.edu.pl/}} is divided into 116 subfields. 

OGLE~III provides data for 17693 RR~Lyrae stars of type {\em ab} \citep{Soszynski08} and 1849 classical Cepheids \citep{Soszynski09}, which all pulsate in the fundamental mode and are distributed over the entire area of the LMC covered by OGLE. For each star the period (with individual uncertainties) and mean magnitudes in the $V$ and $I$ bands are provided. The mean magnitudes have an uncertainty of 0.07~mag. For the $I$ band measurements the resulting amplitude of the lightcurve is given. Furthermore, the OGLE~III collaboration calculates the parameters $R_{21}$ and $R_{31}$, which represent the skewness, as well as $\phi_{21}$ and $\phi_{31}$, corresponding to the acuteness of the lightcurve \citep{Stellingwerf87a}, via Fourier decomposition. 

%

\section{Distance measurements}
\label{distance}

We use RR~Lyrae stars and Cepheids to determine the distances of the old and the young population of the LMC, respectively. For the old population RR~Lyrae stars, which have ages of at least 9~Gyr \citep[e.g.,][]{Sarajedini06}, provide insights into the early stages of the evolutionary history of the galaxy. The Cepheids trace a much more recent time period covering an age range of approximately 30 to 300~Myr \citep[e.g.,][]{Grebel98, Luck03}.

\subsection{RR~Lyrae}

The absolute luminosity $M_V$ of the RR~Lyrae stars depends only on their metallicity. In Paper~II we present individual photometric metallicity estimates for 16949 RR~Lyrae stars. These values on the metallicity scale of \citet{Zinn84} are used together with the widely used relation by \citet{Clementini03} to calculate absolute magnitudes for each single star according to

\begin{equation}
M_V = 0.84 + (0.217 \pm 0.047)~\mathrm{[Fe/H]}
\label{absolute_magnitude}
\end{equation}

We tested various equations for the absolute magnitude of RR~Lyrae stars with quadratic metallicity terms of \citet{Catelan04} and of \citet{Sandage06}. The quadratic equations always lead to slightly higher absolute magnitudes than the relation of \citet{Clementini03}. For the relation of \citet[][their equation 8]{Catelan04} a median difference of $+0.08$~mag is found, while the median difference for \citet[][their equation 7]{Sandage06b} is below $+0.01$~mag. These differences are small and we decide to use the relation of \citet{Clementini03}. 

The absolute magnitude together with the mean observed magnitudes and a reddening correction (see Section~\ref{reddening}) lead to the true distance modulus of each star. This permits us to investigate the three-dimensional distribution of the old stellar population in the LMC as traced by its numerous RR~Lyrae stars. 

\subsection{Cepheids}

Cepheids follow a well-known period-luminosity relation. We adopt the relations derived by \citet{Sandage04a}. For LMC Cepheids these authors found that there are two different sets of relations depending on the length of the period. For periods \textit{shorter} than 10 days they found 

\begin{eqnarray}
M_V = -(2.963 \pm 0.056)\log P - (1.335 \pm 0.036) \label{Cepheid_M_V_s_10} \\
M_I = -(3.099 \pm 0.038)\log P - (1.846 \pm 0.024) \label{Cepheid_M_I_s_10}
\end{eqnarray}
and for periods \textit{longer} than 10 days
\begin{eqnarray}
M_V = -(2.567 \pm 0.102)\log P - (1.634 \pm 0.135) \label{Cepheid_M_V_l_10} \\
M_I = -(2.822 \pm 0.084)\log P - (2.084 \pm 0.111) \label{Cepheid_M_I_l_10}
\end{eqnarray}
where $P$ is the period of the Cepheid in days. 

We apply the appropriate relations to all available Cepheids in the OGLE~III archive and calculate the apparent distance modulus using the mean magnitudes provided by OGLE. In the next step, we correct these apparent distance moduli for extinction effects.

\subsection{Reddening correction}
\label{reddening}

In \citet[hereafter Paper~I]{Haschke11_reddening} we used two different methods to obtain reddening maps of the MCs. One of these methods uses an area-averaged reddening, whereas the other method provides reddening estimates for each individual tracer star. Here we apply both methods to correct the distances calculated above for extinction effects. 

\subsubsection{Area-averaged reddening correction:\\ The red clump method}

RC stars in the LMC are utilized in Paper~I to calculate the reddening on scales of a few arcminutes. Depending on the metallicity of the stars, the RC is situated at a well-defined mean color in the color-magnitude diagram \citep{Girardi01}. When the metallicity is known, the difference between the measured mean color of the RC and the theoretically predicted color provides the mean reddening towards this part on the sky \citep{Wozniak96}. 

In Paper~I we subdivide the OGLE~III area into a grid of smaller fields each of which contains at least a few hundred RC stars from the OGLE~III photometric catalog \citep{Udalski08a}. The size of these subfields is adjusted to the different RC star densities across the area covered by OGLE~III and varies from $4.5 \times 4.5$~arcmin to $36 \times 36$~arcmin. For each subfield the mean color is computed and compared to the theoretical, unreddened color. We adopt the value found by \citet{Olsen02} for the unreddened color of the RC of the LMC. For each position of the 16949 RR~Lyrae stars and 1849 Cepheids the appropriate reddening of its subfield is applied. Using the relations by \citet{Schlegel98} between total and selective extinction in different bands

\begin{eqnarray}
A_V = 3.24(E(V-I)/1.4) \label{reddening_V} \\
A_I = 1.96(E(V-I)/1.4) \label{reddening_I}
\end{eqnarray}
we correct the distance modulus, determined above for every star, for the mean line-of-sight extinction at that star's location on the sky. 

\subsubsection{Individual reddening corrections:\\ Intrinsic colors of variable stars}
\label{reddening_color}

\subsubsection*{RR~Lyrae}

Individual reddening estimates are calculated in Paper~I for each of the 12675 RR~Lyrae stars with a mean magnitude in the $V$ and $I$ band in the OGLE~III LMC database. Differences between the color of the observed mean magnitudes and the theoretically determined absolute magnitudes are computed taking our photometric metallicities into
account. 

In Paper~I we use the relations of \citet{Catelan04} for the absolute magnitudes. They are similar to Equation~\ref{absolute_magnitude} of \citet{Clementini03}, but in addition provide consistent relations for different photometric passbands. In Paper~I we derive the predicted absolute magnitude in the $V$ and $I$ bands and define an ``absolute color'', $(V-I)_0$. The observed apparent color $(v-i)$ of the RR~Lyrae stars is calculated by subtracting the apparent mean $i$ magnitude from the apparent mean $v$ magnitude given by OGLE~III. The difference $E(V-I) = (v-i) - (V-I)_0$ of these two colors yields the reddening. 

We use the Equations~\ref{reddening_V} and \ref{reddening_I} to calculate extinctions for 12675 RR~Lyrae stars and correct their apparent distance moduli accordingly. 

\subsubsection*{Cepheids}

The basic approach of individual reddening estimates is used also to calculate Cepheid reddenings. We apply the equations \ref{Cepheid_M_V_s_10}, \ref{Cepheid_M_I_s_10},
\ref{Cepheid_M_V_l_10}, and \ref{Cepheid_M_I_l_10} to calculate $(V-I)_0$ for each Cepheid based on its period. With the mean magnitudes from the OGLE~III catalog, the observed apparent color $(v-i)$ is computed. We then derive the reddening from the difference of the apparent and absolute colors and use it in order to calculate the absolute distance moduli of the 1849 Cepheids. 

%

\section{Star densities in the OGLE~III field}
\label{Density_of_OGLEIII}

\begin{figure}
\centering 
 \includegraphics[width=0.47\textwidth]{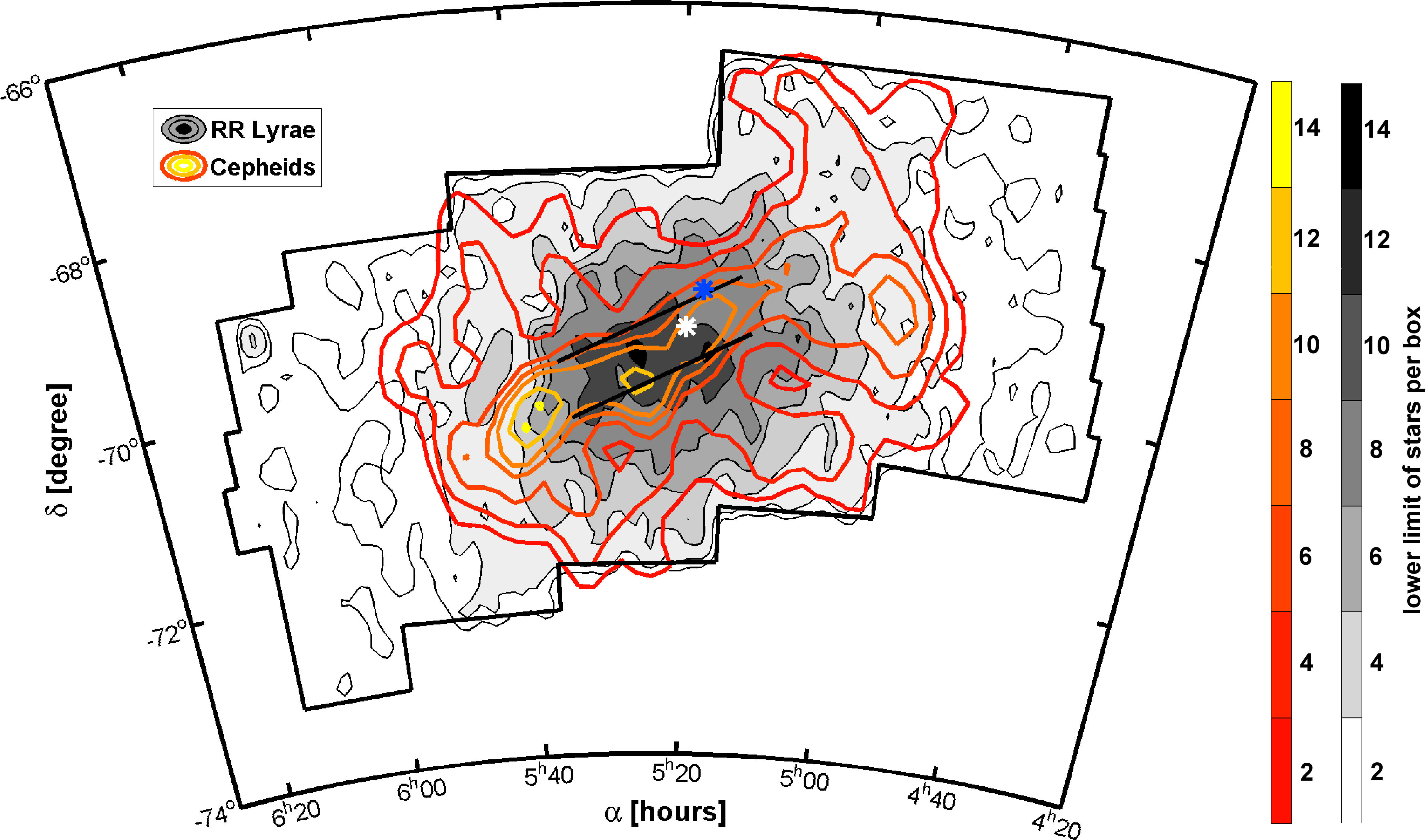}
 \caption{Projected densities of RR~Lyrae stars (filled grey contours) and Cepheids (colored contours) as a function of right ascension, $\alpha$ (J2000), and declination, $\delta$ (J2000). The highest densities of RR~Lyrae stars are found in the bar region. Further away from the bar the RR~Lyrae distribution becomes much more
circular. The high density regions of the Cepheid distribution are prolate and trace the bar. The thick black polygon delineates the boundaries of the OGLE~III region. The optical center of the LMC found by \citet{Vaucouleurs72} is marked with a white asterisk. The kinematical center from H~{\sc i} observations of \citet{Kim98b} is marked by a blue asterisk. The box sizes of the evaluated fields are listed in Table~\ref{table_bins_RC_contour}.} 
 \label{RRL_Cep_RADEC_colourext}
\end{figure}

The OGLE~III data cover nearly 40 square degrees of the LMC. Figure~\ref{RRL_Cep_RADEC_colourext} shows the distribution of the RR~Lyrae stars and the Cepheids located in this area. The number of RR~Lyrae stars is counted in fields of $0.25^{\circ} \times 0.125^{\circ}$ in right ascension and declination, respectively. We find the center of the RR~Lyrae distribution at $\alpha = 5^h26^m$ and $\delta = -69^{\circ}75'$. Due to the much lower density of the Cepheids their distribution is evaluated in larger fields of $0.5^{\circ} \times 0.25^{\circ}$ in right ascension and declination, respectively (Table~\ref{table_bins_RC_contour}). We then smooth the resulting density distributions using a Gaussian kernel. To illustrate the location of the bar region we adopt its approximate location from \citet{Zhao00} (marked with the two black diagonal lines in Figure~\ref{RRL_Cep_RADEC_colourext}). 

For the RR~Lyrae stars the highest density of stars is found along the bar. The highest density of RR~Lyrae stars does not coincide with the kinematical or optical center of the LMC. With increasing distance from the bin of highest density and with decreasing density of stars the distribution becomes more ellipsoidal. Unfortunately the OGLE~III data do not cover the low density outskirts of the LMC in the southern and northern directions. 

For the Cepheids we find a different picture than for the old RR~Lyrae population. The distribution of stars is elongated and the highest density of stars is again found along the bar region. The Cepheids show different locations of enhanced density. They are formed at various locations with varying intensities reflecting regions of enhanced star formation activity during the past 30 to 300~Myr \citep[e.g.,][]{Grebel98}. Neither the RR~Lyrae nor the Cepheid density maximum coincide with the optical center of the LMC found by \citet{Vaucouleurs72} or with the kinematical center found by \citet{Kim98b} (see Figure~\ref{RRL_Cep_RADEC_colourext}). 

Using data from the Two Micron All Sky Survey (2MASS) \citep{Skrutskie06} and Deep Near-Infrared Southern Sky Survey (DENIS) \citep{Epchtein97} the distribution of RGB and AGB stars in the LMC was investigated by \citet{Marel01b}. \citet{Cioni06} found average ages of 5--6~Gyr for the AGB stars. \citet{Salaris05} obtained average ages of about 4~Gyr for the RGB stars of the LMC. In Figure~2 of \citet{Marel01b} the density distribution of these intermediate-age stars is shown. The bar region is clearly visible and has the highest stellar density. Towards the outskirts of the LMC the density distribution becomes first rounded, and then elongated in the north-south direction, approximately perpendicular to the bar. These outer parts of the density map of \citet{Marel01b} are not covered by the OGLE~III area and can therefore not be compared with the density distribution of Cepheids and RR~Lyrae stars that we obtain in our present paper. 

The distribution of RR~Lyrae stars closely resembles that of the intermediate-age (AGB/RGB) star density distribution in the central region of the LMC while the Cepheid distribution shows more substructure. With younger age of the stellar populations (i.e., for Cepheids) the high density regions become increasingly elongated along the LMC bar, extending further in eastern and western direction than the main body of the bar traced by the RR~Lyrae stars \citep[compare also the Boxes 7 and 9 in Figure~5 of][]{Bastian09}.

\begin{table}
\begin{center}
\caption{Binsizes of the fields evaluated to obtain the densities of
RR~Lyrae and Cepheid stars in the LMC. The grid of the boxes defined
here is used in Figure \ref{RRL_Cep_RADEC_colourext},
\ref{RRL_Cep_dist_RCext}, and \ref{RRL_Cep_dist_colourext} to show the
distribution of young and old stars in the LMC.
\label{table_bins_RC_contour} } 
\begin{tabular}{c c c} 
\tableline\tableline
 & RR~Lyrae & Cepheids \\ 
\tableline 
$\alpha$ bin [degree] & 0.25 & 0.5 \\ 
$\delta$ bin [degree] & 0.125 & 0.25 \\ 
distance bin [kpc] & 0.25 & 0.5 \\ 
\tableline  
\end{tabular}
\end{center}
\end{table}

%

\section{Three dimensional maps}
\label{3D}

The distances of RR~Lyrae and Cepheids are calculated with the equations described in Section~\ref{distance}. We obtain four sets of maps, two for each the young and the old stars when applying either average RC or individual reddening corrections.

\subsection{Maps corrected with RC reddening}
\label{corrected_RC}

\begin{figure}
\centering 
 \includegraphics[width=0.47\textwidth]{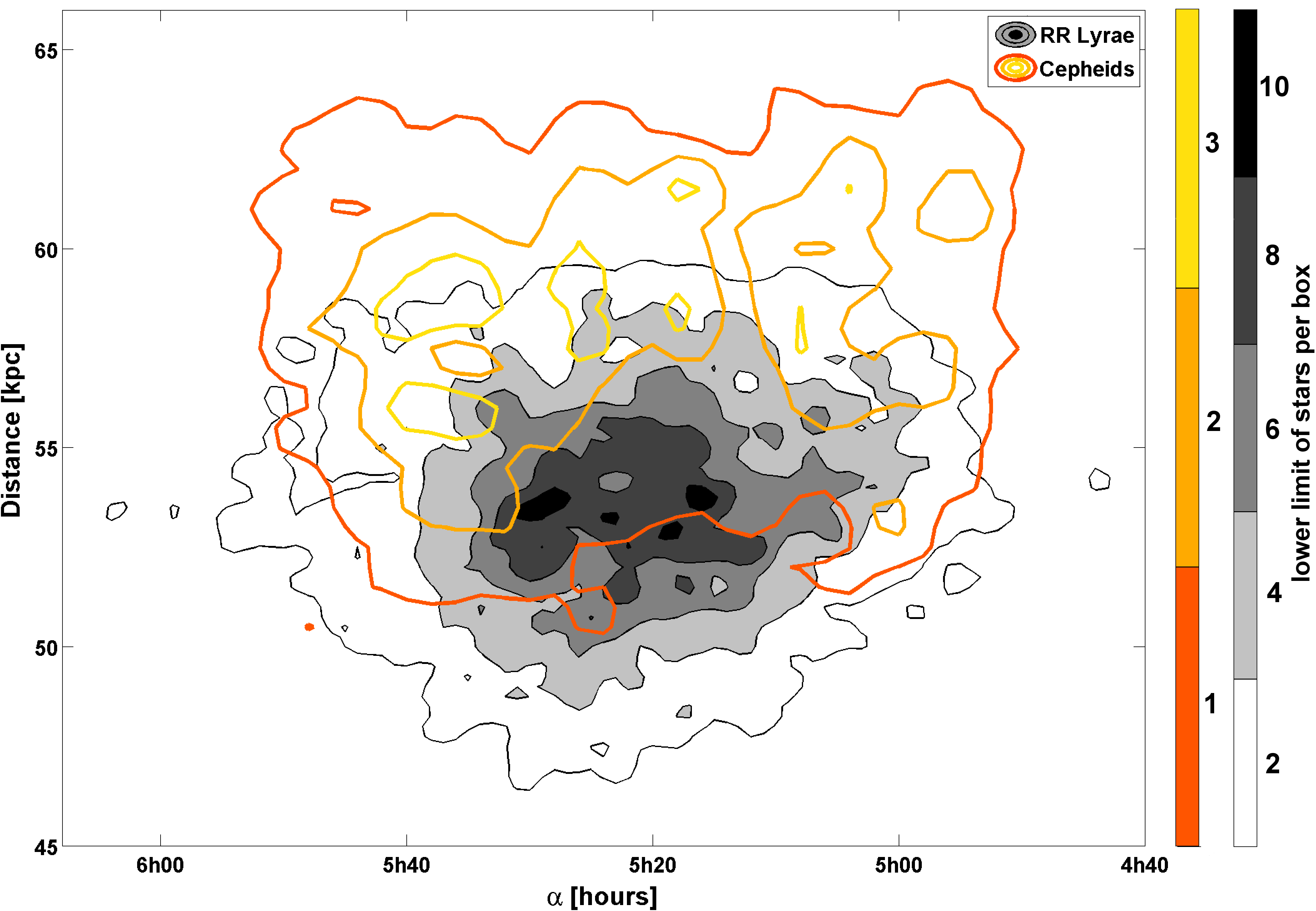}
\vspace{\floatsep} 
 \includegraphics[width=0.47\textwidth]{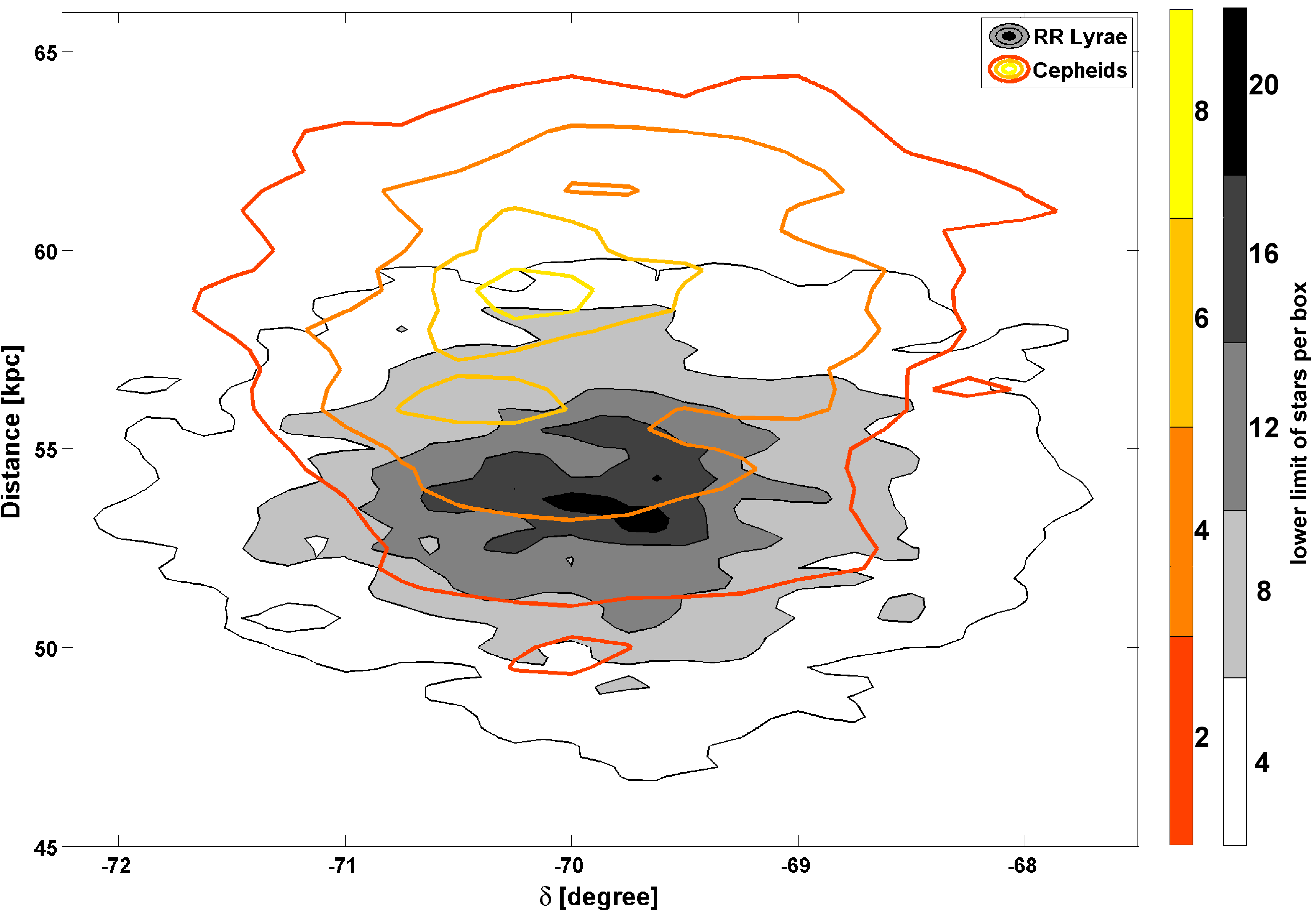}
\caption{Stellar densities of RR~Lyrae stars (filled grey contours) and Cepheids (colored contours) as a function of distance and right ascension in the upper panel and as a function of distance and declination in the lower panel. All distances are extinction-corrected using the mean RC reddening values. The main concentration of the
RR~Lyrae is closer and more centrally concentrated than that of the Cepheids. The box sizes of the evaluated fields are listed in Table~\ref{table_bins_RC_contour}.} 
\label{RRL_Cep_dist_RCext}
\end{figure}

In this subsection we apply the reddening estimates derived with the RC method to correct the distances of the RR~Lyrae stars and Cepheids obtained in Section~\ref{distance}. In the upper panel of Figure~\ref{RRL_Cep_dist_RCext} we plot the density distribution of RR~Lyrae stars (grey contours) and Cepheids (colored contours) in right ascension versus distance. In the lower panel of Figure~\ref{RRL_Cep_dist_RCext} these density distributions are plotted as a function of declination vs.\ distance. This implies a view from above the LMC in the upper panel and from the side in the lower panel. The field sizes within which the density of stars is evaluated are listed in Table~\ref{table_bins_RC_contour}. The contour plots are smoothed with a Gaussian kernel, which uses $3 \times 3$ bins and a width of 1 bin to reduce the variance on very small scales. For the uncertainties of the density contours we assume Poisson noise. The noise is usually smaller than the binsizes. 

\begin{table*}
\begin{center}
\caption{Recent distance estimates of the LMC using RR~Lyrae stars or Cepheids. In the last line the mean distance to the LMC found by \citet{Alves04} using many different distance indicators is stated.} 
\label{distance_table} 
\begin{tabular}{c c c l} 
\tableline \tableline
type of indicator & $(m-M)_{mean}$ & $\sigma$ & reference \\ \tableline 
Cepheids & 18.35 & 0.13 & \citet{Luri98} \\ 
Cepheids & 18.60 & 0.11 & \citet{Groenewegen00} \\ 
Cepheids & 18.56 & 0.10 & \citet{Gieren05} \\
Cepheids & 18.56 & 0.03 & \citet{Benedetto08} \\
Cepheids & 18.48 & 0.04 & \citet{Ngeow08} \\ 
Cepheids & 18.45 & 0.04 & \citet{Storm11}\\
Cepheids & 18.85 & 0.08 & this work - area averaged reddening \\ 
Cepheids & 18.65 & 0.07 & this work - individual reddening \\ 
RR~Lyrae & 18.45 & 0.07 & \citet{Clementini03} \\ 
RR~Lyrae & 18.43 & 0.16 & \citet{Alcock04} \\
RR~Lyrae & 18.48 & 0.08 & \citet{Borissova04} \\
RR~Lyrae & 18.54 & 0.09 & \citet{Marconi05} \\
RR~Lyrae & 18.44 & 0.11 & \citet{Catelan08} \\ 
RR~Lyrae & 18.58 & 0.11 & \citet{Szewczyk08} \\ 
RR~Lyrae & 18.53 & 0.13 & \citet{Borissova09} \\ 
RR~Lyrae & 18.60 & 0.17 & this work - area averaged reddening \\ 
RR~Lyrae & 18.62 & 0.13 & this work - individual reddening \\ 
\tableline 
multiple indicator mean & 18.50 & 0.04 & \citet{Alves04} \\ 
\tableline
\end{tabular}
\end{center}
\end{table*}

The spatial distribution of the RR~Lyrae and Cepheids form distinct groups whose centroids do not coincide. Without correcting for the inclination, computing the median distance of the RR~Lyrae stars leads to $D_{\mathrm{RRL/median}} = 52.7 \pm 3.9$~kpc ($(m-M)_0 = 18.60 \pm 0.17$). For the Cepheids we find a distance of $D_{\mathrm{Cep/median}} = 58.8 \pm 2.2$~kpc ($(m-M)_0 = 18.85 \pm 0.08$). 

The uncertainties are computed using error propagation adopting a mean magnitude error of 0.07~mag, as stated by the OGLE collaboration, the mean extinction uncertainty of 0.08~mag from Paper~I and a metallicity uncertainty for the RR~Lyrae stars of 0.23~dex (Paper~II). The uncertainty of the period is so small that it can be neglected. Taking the errors into account the resulting median distances of the old and young population agree within 2$\sigma$. 

By deprojecting the spherical coordinates into a Cartesian coordinate system \citep{Weinberg01}, using the optical center of \citet{Vaucouleurs72} and the structural parameters determined in Section \ref{3D_structure}, the median distances change slightly to $D_{\mathrm{RRL/median}} = 54.4 \pm 3.9$~kpc and $D_{\mathrm{Cep/median}} = 58.6 \pm 2.2$~kpc. These values are very similar to the ones without deprojection. We therefore use the mean distances as measured without correcting for the inclination angle.

Our RR~Lyrae distance is in agreement with the literature values. The Cepheid distance differs by more than 3$\sigma$ from the mean values of the literature (see Table~\ref{distance_table}). 

\subsection{Maps corrected with individual reddening}

In the previous section we used reddening values based on the color of the RC (see \citetalias{Haschke11_reddening} for details). Instead we can also compute the reddening for each of our Cepheid or RR~Lyrae stars individually as described in the Section~\ref{reddening}, which is a more accurate method that does not average over an area and that does not use other stellar populations (RC stars) as tracers of the reddening of much younger or much older populations. For a more detailed discussion of population- or temperature-dependent reddening see \citet{Grebel95} and \citet{Zaritsky99}. 

\begin{figure}
\centering 
 \includegraphics[width=0.47\textwidth]{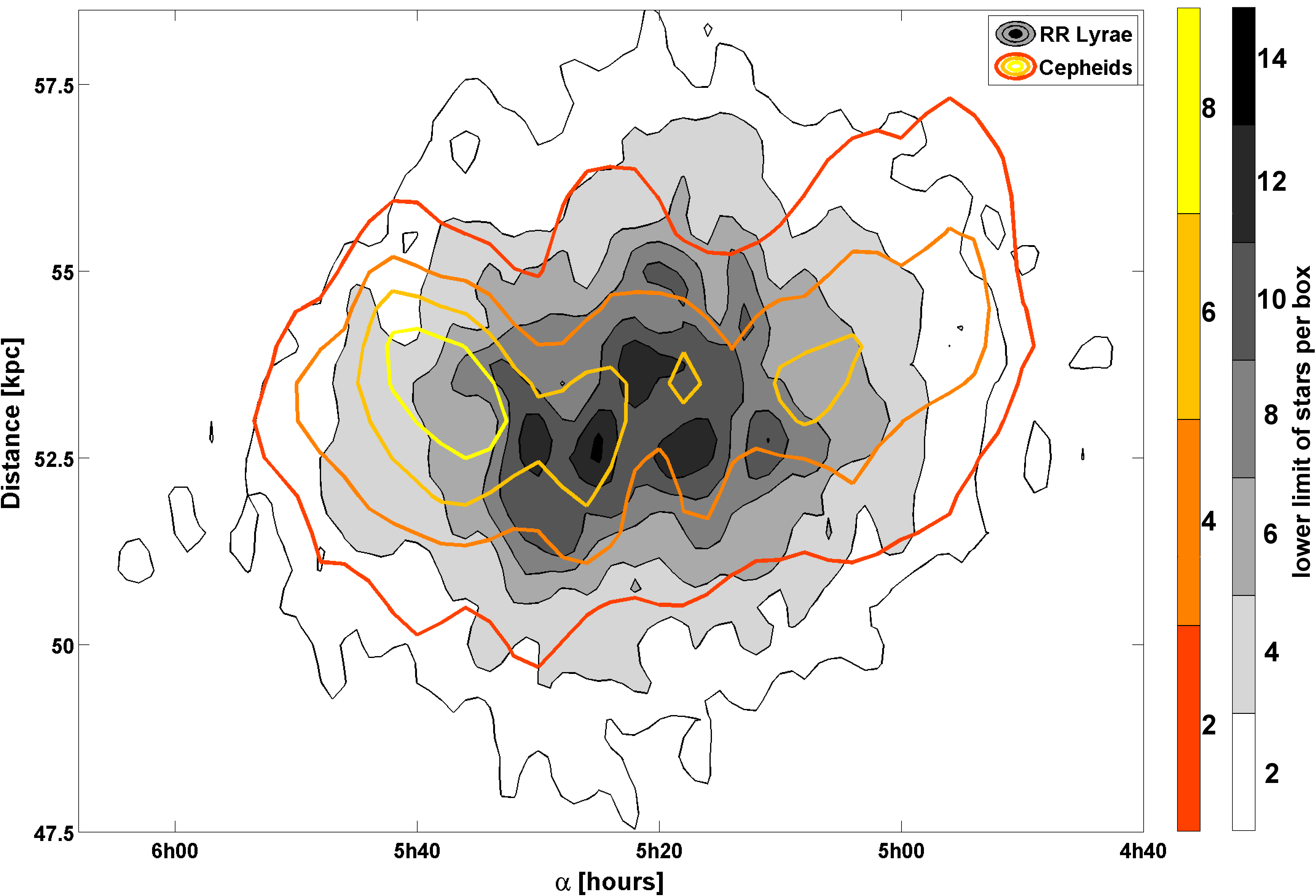}
\vspace{\floatsep}
 \includegraphics[width=0.47\textwidth]{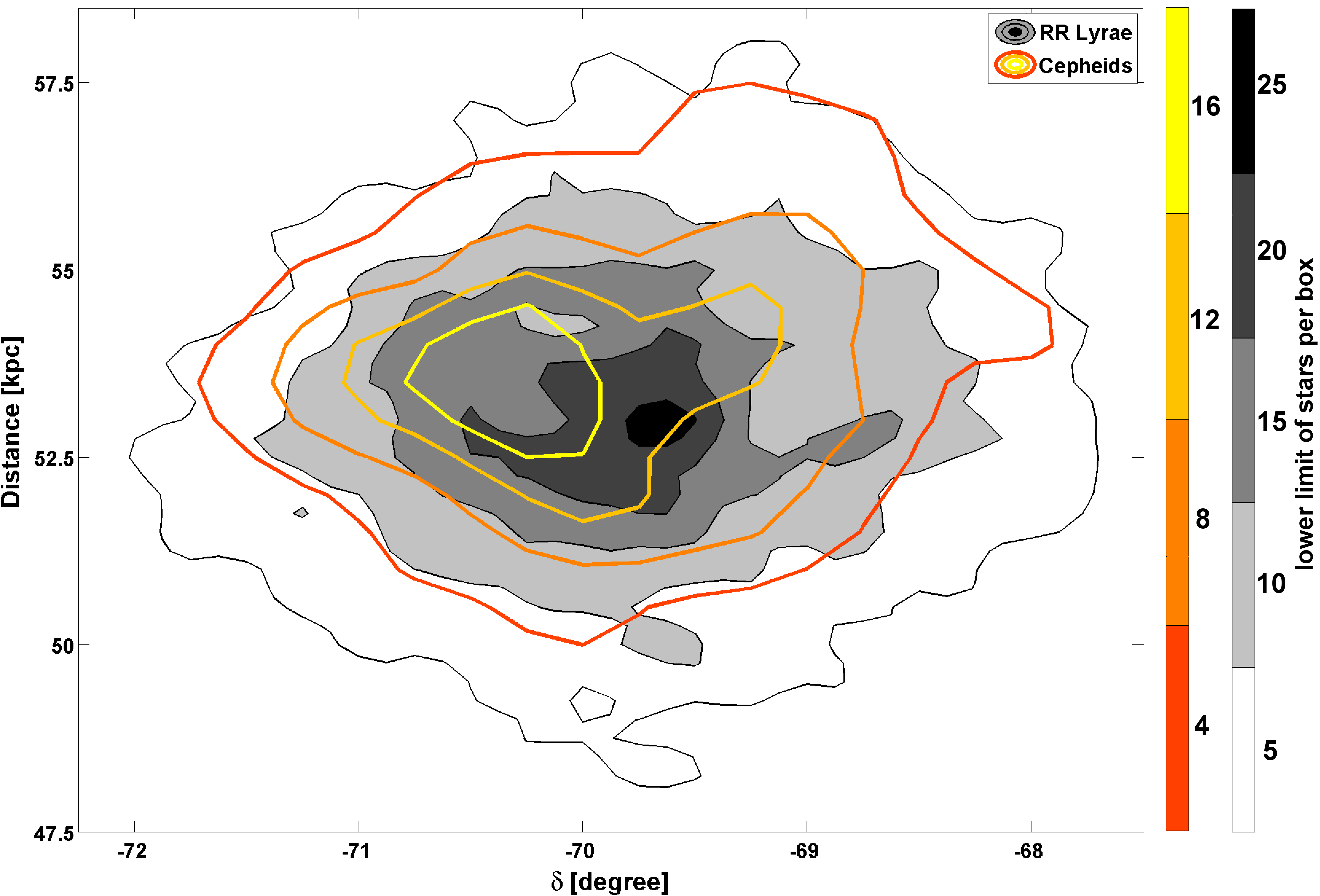}
\caption{Stellar densities of RR~Lyrae stars (filled grey contours) and Cepheids (colored contours) are shown as a function of distance and right ascension in the upper panel and of distance and declination in the lower one. All distances are extinction-corrected using the individual, intrinsic color reddening values. The Cepheids agree very well with the RR~Lyrae distances within the errors. The Cepheids are not as centrally concentrated as the RR~Lyrae and trace the elongated distribution of the bar. The box sizes of the evaluated fields are listed in Table~\ref{table_bins_RC_contour}.} 
 \label{RRL_Cep_dist_colourext} 
\end{figure}

The resulting maps (Figure~\ref{RRL_Cep_dist_colourext}) show a reduced depth of the LMC as compared to Figure~\ref{RRL_Cep_dist_RCext}. The computed distances of the Cepheids decrease as compared to Section~\ref{corrected_RC} (without having corrected for the intrinsic shape of the LMC), while the RR~Lyrae distances remain similar to those obtained with the RC reddening approach. We find a median distance for all RR~Lyrae stars of $D_{\mathrm{RR~Lyrae/median}} = 53.1 \pm 3.2$~kpc ($(m-M)_0 = 18.62\pm 0.13$) and for the Cepheids of $D_{\mathrm{Cep/median}} = 53.9 \pm 1.8$~kpc ($(m-M)_0 = 18.65\pm 0.07$). The deprojection of the coordinates, using the same parameters as above, leads to very similar distances of $D_{\mathrm{RR~Lyrae/median}} = 53.4 \pm 3.2$~kpc and $D_{\mathrm{Cep/median}} = 54.0 \pm 1.8$~kpc. Later on we refer only to the non-deprojected overall mean distances, because most of the published mean distance estimates for the LMC have been calculated without a deprojection. To facilitate the comparison with other investigations, we choose to keep the non-deprojected overall mean distances as our main result. 

A mean magnitude error of 0.07~mag is stated by OGLE, a mean extinction error of 0.06~mag is found in \citetalias{Haschke11_reddening} and an uncertainty for the metallicity of 0.23~dex is estimated by \citetalias{Haschke12_MDF}. The resulting distance errors are calculated using error propagation. The error of the period is so small that it can be neglected. 

The change in the median RR~Lyrae distance of the LMC is small and agrees well within the uncertainties with the previously calculated value from Section~\ref{corrected_RC}. For the Cepheids the change in distance calculated with the individual extinction corrections and with the area-averaged reddening values of the RC method is greater than the uncertainties of both determinations together. Cepheids undergo considerable mass loss, which may lead to an accumulation of circumstellar dust around these stars \citep{Barmby11}. This may result in additional differential reddening not accounted for by the RC maps. Moreover Cepheids are located in or close to recent star-forming regions with potentially higher dust and gas content. This may lead to fluctuations and variations on scales that are not resolved by the RC method \citepalias[see][]{Haschke11_reddening}. Also other photometric studies show that the applicable extinction changes with the stellar population considered \citep[e.g.,][]{Zaritsky02, Zaritsky04}.

\subsection{Cepheid clustering as a test of the de-reddening method} 
\label{reddening_test}

\begin{figure}
\centering 
 \includegraphics[width=0.47\textwidth]{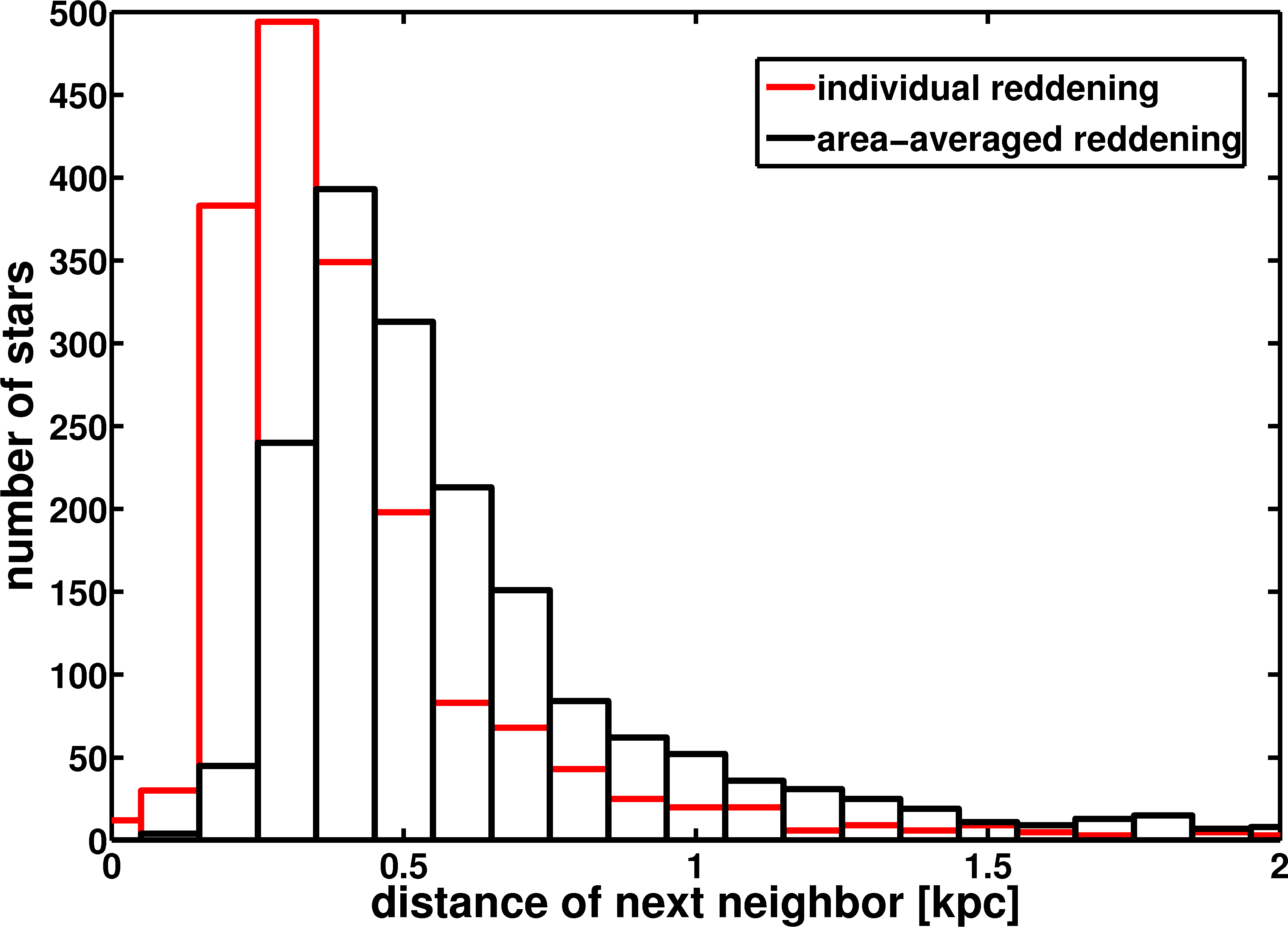}
 \caption{Histogram distribution of the distances to the fifth nearest Cepheid neighbor of each Cepheid in our sample. The individually dereddened Cepheids (grey histograms or red histograms in the on-line version) have smaller mean distances from their nearest neighbors than the Cepheids dereddened via the area-averaged method
(black histograms).} 
 \label{next_neighbor}
\end{figure}

Star formation usually occurs clustered \citep{Lada03}, leading to soon dissolving, loose associations, open clusters, or more tightly bound populous clusters that may survive for many Gyr. In an analysis of the distribution of young, age-dated populations in the LMC, \citet{Bastian09} found average dissolution time scales of the order of $\sim 175$ Myr. 

Considering that Cepheids cover an approximate age range of some 30 to 300 Myr, we may therefore expect that their spatial distribution should still reflect some of this clustering, as many of the Cepheids would not yet have moved too far from their birth places. Therefore we expect the mean distance of the Cepheids to their next Cepheid
neighbors to be small. Projected on the sky plane we can directly observe this close proximity of stars. Regarding the three-dimensional distribution of Cepheids, we expect that distances calculated with a superior reddening method to reflect this by a smaller mean distance between adjacent Cepheids along the z-axis as well.

For each Cepheid we calculate the three-dimensional distances to all other Cepheids present in the sample. Then we determine the distance to the fifth and tenth closest neighbor of each Cepheid. We find that the individually dereddened Cepheids have smaller mean distances to their next neighbors than the Cepheids with an area-averaged dereddening (Figure~\ref{next_neighbor}). For the fifth nearest neighbors a median distance of $0.35$~kpc and $0.52$~kpc is found for the two de-reddening approaches, respectively. The tenth nearest neighbors have a median distance of $0.50$~kpc and $0.75$~kpc, respectively. Apart from the arguments in favor of individual de-reddening discussed in Paper~I and in the previous subsection, the smaller mean distances between the Cepheids provide further support for the use of this method as opposed to the area-averaged reddening estimates from RC stars.

\subsection{Similar mean LMC distances for Cepheids and RR Lyrae stars}

Although the mean distances of the young and old population agree now within their uncertainties, the Cepheids are distributed differently than the RR~Lyrae stars. While the RR~Lyrae are densest in the center of the LMC, the highest concentrations of the Cepheids closely follow the entire extent of the bar. The upper panel of Figure~\ref{RRL_Cep_dist_colourext} shows two concentrations of Cepheids towards the eastern and western end of the bar, while in the lower panel a southward shift of the Cepheids is visible. This kind of behavior is typical for irregular galaxies: Younger populations trace the often widely scattered distribution of recent star-forming regions, while older populations show a much more regular spherical distribution \citep[see also, for instance,][]{Zaritsky00, Grebel01, Tikhonov05, Bastian09, Glatt10, Crnojevic11}.

Our mean LMC distance from RR~Lyrae stars is, within 1~$\sigma$, in good agreement with the values found in the literature for LMC RR~Lyrae stars as shown in Table~\ref{distance_table} as well as with the mean distance of many different distance tracers given in the review by \citet{Alves04}. For the Cepheid results from the literature the uncertainties of the distances are typically smaller than for the RR~Lyrae distances. Within 1~$\sigma$, our new Cepheid distance modulus does not agree with most of the literature values or the mean value by \citet{Alves04}. However, our mean Cepheid distance agrees with the Cepheid distances of \citet{Groenewegen00} using Hipparcos data. 

No indication for a differing ``long'' and ``short'' distance scale is found in our work when using the individual reddening corrections for Cepheids and RR Lyrae stars. This is in contrast to the more discrepant distances resulting when using the RC reddening correction for Cepheids and RR~Lyrae stars. Clearly our new extinction measurements play a major role for the resulting distances. 

We note that any distance determination, whether for Cepheids, RR Lyrae stars, or other distance indicators, does critically depend on the adopted zero-point calibrations. As reviewed by \citet{Walker11}, these calibrations have much improved over recent years, reducing the discrepancies between different indicators. As Walker points out, one of the primary remaining contributing uncertainties is the reddening (which we are addressing by determining individual reddening values for each star). Other uncertainties come from, e.g., possibly unrecognized metallicity effects and remaining unsolved problems in understanding post-main-sequence evolution. Future breakthroughs may be expected from major stellar surveys focusing on infrared wavelengths where reddening and metallicity effects are reduced \citep[e.g.,][]{Storm11, Ripepi12} and the parallax measurements by the forthcoming Gaia satellite mission of the European Space Agency \citep[e.g.,][and references therein]{Eyer12}.

%

\section{Three dimensional structure}
\label{3D_structure}

\subsection{The LMC in slices}

Our data permit us also to obtain more information about the location and orientation of the young and old population of the LMC. In this subsection we slice the LMC by using only stars in distinct distance bins to reveal more information about the LMC's structural properties.

\subsubsection*{RR~Lyrae stars}

\begin{figure*}
\centering 
 \includegraphics[width=1\textwidth]{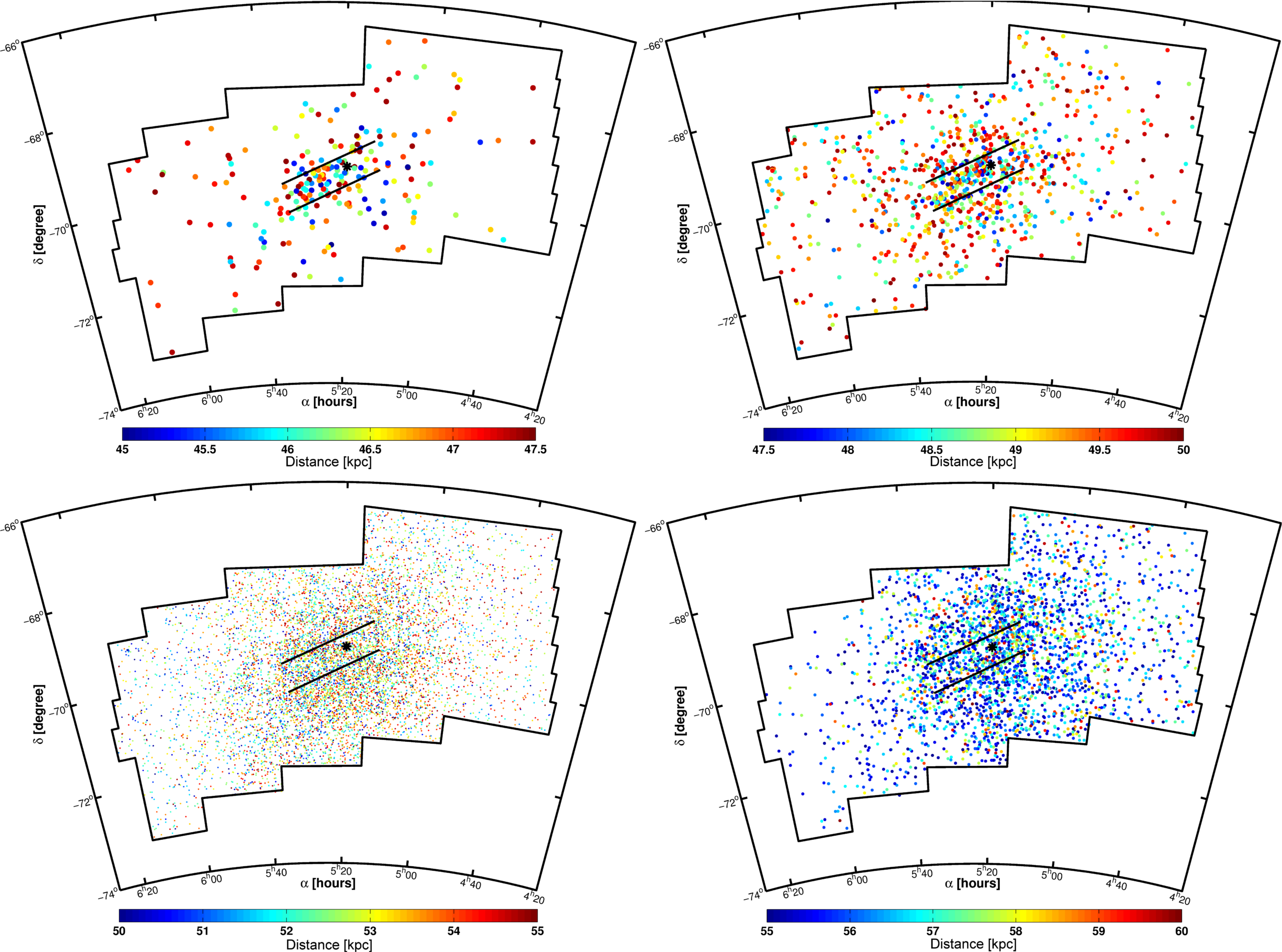}
\caption{Spatial position of the RR~Lyrae stars for four different distance ranges. Each star is color-coded with its distance as quantified in the color bar at the bottom of each panel. Each panel shows a different distance regime of the LMC, altogether covering line-of-sight distances of 45 to 60~kpc. In the closest regions the bar is very well visible, while it disappears as a distinct overdensity in the farther regions. The eastern region of the LMC is closer than the western one. The individual distance uncertainties are of the order of $8\%$ of the calculated distance. The black polygon shows the boundaries of the OGLE~III field and the two black diagonal lines show the approximate location of the bar. The black star represents the optical center of the LMC found by \citet{Vaucouleurs72}. The upper left panel contains 221~stars, the upper right panel 893~stars, the lower left panel 8151~stars and the lower right one 2916~stars.} 
 \label{RRL_dist_45_60} 
\end{figure*}

In Figure~\ref{RRL_dist_45_60} we show the projected distribution of 12675 RR~Lyrae stars in four different distance bins. The upper two panels show the closest parts of the LMC with RR~Lyrae stars in a distance range between 45 to 47.5~kpc (left panel) and 47.5 to 50~kpc (right panel). These bins are of the same order as the typical distance uncertainties of individual RR~Lyrae stars at this distance, i.e., about 3~kpc. Figure~\ref{RRL_dist_45_60} demonstrates that the stars are highly concentrated in the central region of the LMC. The upper left panel indicates that the closest stars are mostly located in the bar, and that more stars are located in the eastern part of the LMC at these close distances. The total number of stars in the two distance bins considered above is 1114 (9\% of the sample). The density east of the center of the LMC is 50\% larger than in the western part in this distance range. 

In the lower left panel of Figure~\ref{RRL_dist_45_60} the distance range between 50 to 55~kpc is displayed. We increased the bin size because the individual uncertainties grow with the distance. Overall this distance slice contains the highest number of stars -- 8151 RR~Lyrae stars (67\% of the sample). The central region of the LMC is particularly densely populated. The bar does not stand out as a distinct feature in this distance slice, while star counts reveal that the number distribution in the western and eastern parts is approximately equal. The most distant set of RR~Lyrae stars is shown in the lower right panel with distances between 55 and 60~kpc and contains 2916 stars (24\% of the sample). The central region is less densely populated than in the closer bins, and as before the bar is no longer visible. For the western parts of the LMC we find 60\% more stars than in the eastern parts at this distance, clear evidence for a tilt of the LMC.

\begin{figure} 
\centering
\includegraphics[width=0.47\textwidth]{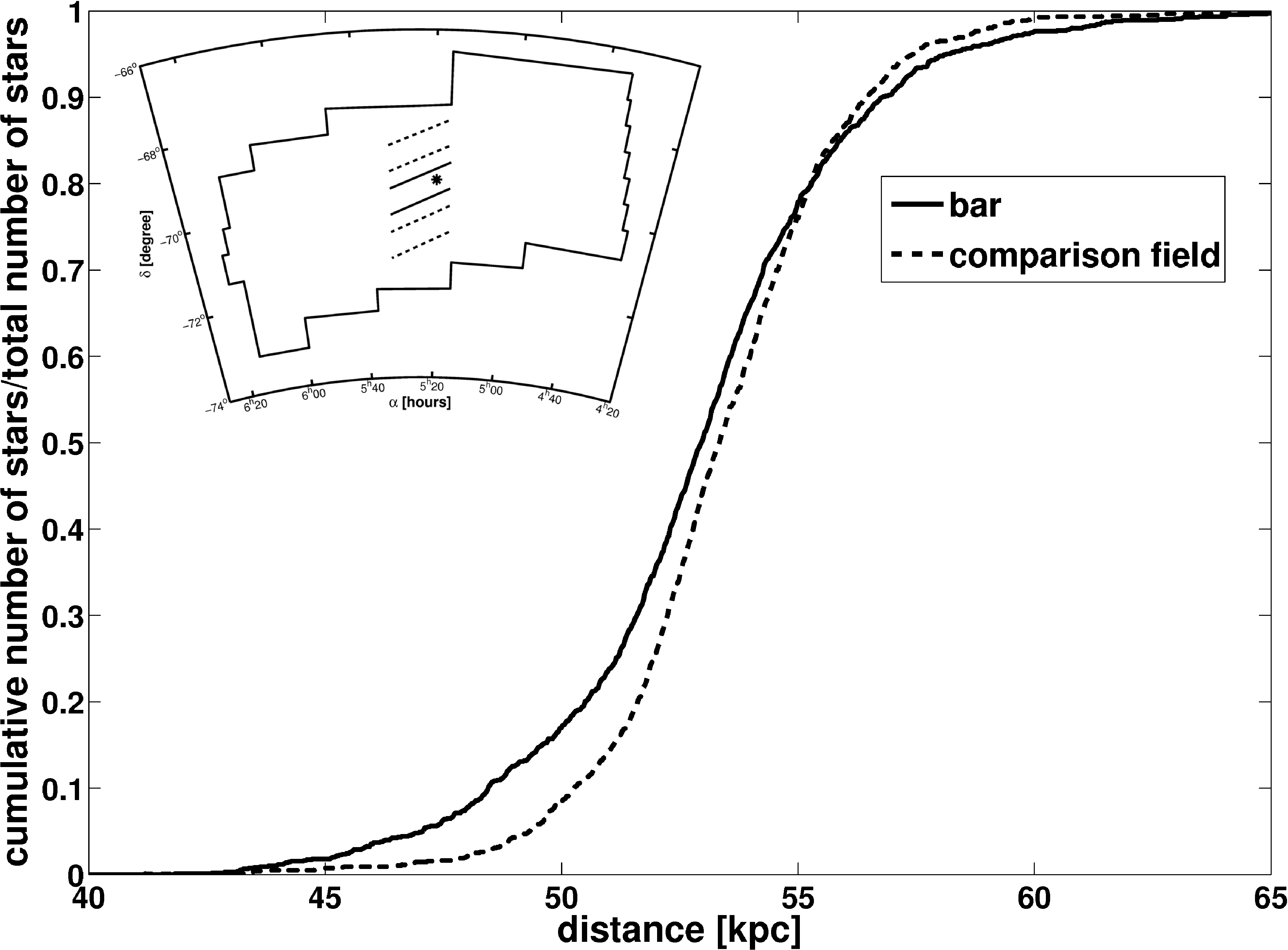} 
\caption{Comparison of the normalized cumulative distribution of the line-of-sight distances of RR~Lyrae stars of the bar region (solid line) with the distribution of the sum of two similar fields located above and below the bar (dashed line). The inset on the upper left shows the location of our three fields superimposed on the OGLE footprint of the LMC. The cumulative distribution of the stars in the region of the bar shows a higher stellar density at closer distances than in the comparison fields.} 
\label{cum_dist}
\end{figure}

In order to further quantify the location of the bar with respect to the disk, we compare the cumulative distribution of stars along the line of sight to the bar with the adjacent regions. In Figure~\ref{RRL_dist_45_60} we draw the approximate northern and southern boundaries of the bar of the LMC. The dimensions and orientation are taken from \citet{Zhao00} and \citet{Mancini04} and were optimized by visual inspection. These boundaries define a parallelogram that contains most of the bar of the LMC. We plot a normalized cumulative distribution of the line-of sight distances of the RR~Lyrae stars within these bar boundaries in Figure~\ref{cum_dist}. We compare this distribution with the normalized cumulative distribution of RR~Lyrae stars in two fields of the same size shifted by $0.6^\circ$ above and below the bar (as marked in the inset in Figure~\ref{cum_dist}). In order to avoid being influenced by the inclination of the disk of the LMC or by warps we add up the data of the two fields outside of the bar region. 

We find the stars in the bar region to stand out as an overdensity largely in front of the main body of the disk. The overdensity can be traced across line-of-sight distances ranging from about 45~kpc to about 53~kpc. The dense part of the bar region thus appears to extend across some 8~kpc in diameter along the line of sight, overlapping at its far end with the main body of the disk. We check for the significance of the difference of the two cumulative distributions by applying a Kolmogorov-Smirnov test (K-S test). With more than 99\% confidence we can rule out that the two distributions come from the same continuous distribution. 

\subsubsection*{Cepheids}

\begin{figure*}
\centering 
 \includegraphics[width=1\textwidth]{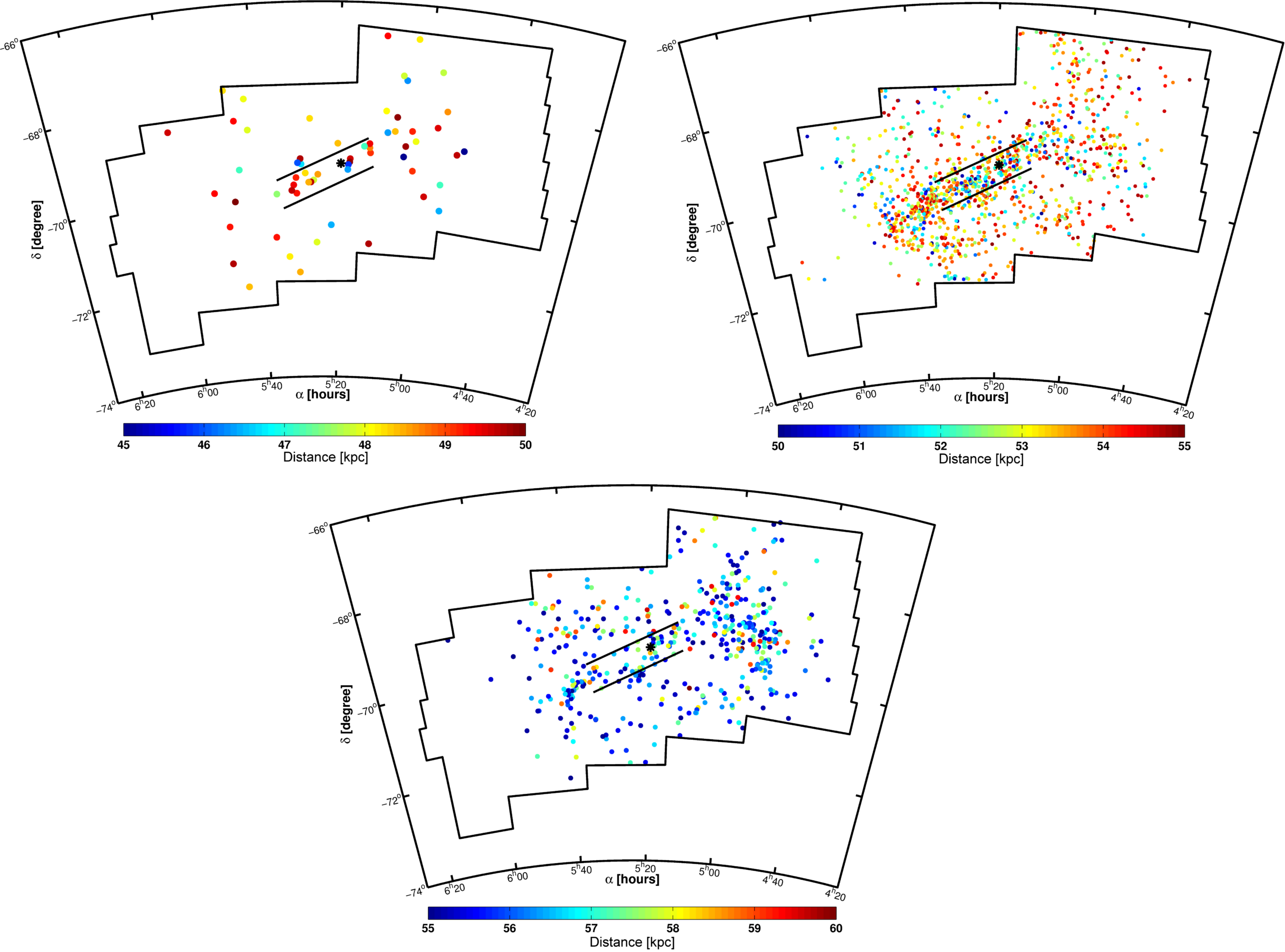}
\caption{Position of Cepheids in the LMC color-coded by their distance in different distance slices. Each panel shows a 5~kpc distance slice, with the upper left panel comprising 62 stars, the upper right panel including 1207 stars and the lower left panel containing 470 stars. Overall, these three panels cover a distance range from 45 to 60~kpc. The bar stands out most clearly in the medium distance slice (upper right panel). The bar extends further to the east and west when traced by Cepheids than for the RR Lyrae stars.} 
 \label{Cep_dist_45_60} 
\end{figure*}

In Figure~\ref{Cep_dist_45_60} the distance distribution of the Cepheids is subdivided into three slices of 5~kpc depth each, covering distances in the range from 45 to 60~kpc. Overall the density of the Cepheids found by OGLE is smaller by a factor of ten compared to the RR~Lyrae stars. 

The closest part of the LMC is represented by a distance slice between 45 and 50~kpc (Figure~\ref{Cep_dist_45_60}, upper left panel). While nearly 10\% of all RR~Lyrae stars are located in this distance range, only 1\% of all Cepheids are found in this slice. Out of these nearly 20\% of the Cepheids are located in the bar region, whose position is marked with the black diagonal lines in Figure~\ref{Cep_dist_45_60}. In the intermediate distance range between 50 and 55~kpc (Figure~\ref{Cep_dist_45_60}, upper right panel) the bar and its extension to the east and to the west is a well visible feature, while the eastern parts of the observed field ($\alpha > 6^h00^m$) are
nearly without any Cepheids. 

In the farthest part of the LMC at distances of 55 to 60~kpc (Figure~\ref{Cep_dist_45_60}, lower panel) the density of Cepheids in the bar has considerably diminished and the panel is dominated by a concentration of stars west of the center. The eastern part of the Cepheid map is devoid of stars beyond $\alpha > 5^h50^m$. 

Thus while the eastern part of the LMC is closer to us, the western part is located further away. We conclude that the overall shape of the disk is similar for the old and young stars in the LMC, in agreement with earlier studies.

\subsection{Inclination angle}

The structural properties of smaller sections of the LMC are investigated by subdividing the OGLE~III field of the LMC in a grid of $0.3^{\circ} \times 0.3^{\circ}$ boxes in right ascension and declination. We calculate the median distance to all stars in these boxes. The inclination angle is computed with a linear fit over all these median distances. 

\begin{table*}
\begin{center}
\caption{A compilation of inclination ($i$) and position angle ($\theta$) values from the recent literature.} 
 \label{inclination_table} 
\begin{tabular}{r c c c c l}
\tableline\tableline 
Type of Stars & $\theta$ [degrees] & $\sigma$ [degrees] & $i$ [degrees] & $\sigma$ [degrees] & Reference \\ 
\tableline 
Cepheids & 151 & 3 & 31 & 1 & \citet{Nikolaev04} \\ 
RC & 114 & 22 & --- & --- & \citet{Subramaniam04} \\ 
RC & 154 & 1 & 23.5 & 0.4 & \citet{Koerwer09} \\ 
Red giants & 122 & 8 & --- & --- & \citet{Marel01a} \\ 
Red giants & --- & --- & 34 & 6 & \citet{Marel01b} \\
Red giants & --- & --- & 29 & --- & \citet{Lah05} \\
RR~Lyrae & 124 & 12 & 31 & 4 & \citet{Subramaniam09a} \\
RR~Lyrae & 112 & --- & --- & --- & \citet{Pejcha09} \\
\tableline 
\end{tabular}
\end{center}
\end{table*}

For the RR~Lyrae stars the western part is $2.5 \pm 0.5$~kpc farther away than the eastern part of the LMC. A similar trend is found for the Cepheids, where the median distance to the western part is $2.7 \pm 0.5$~kpc larger than to the eastern part. These values correspond to an inclination angle of $32^\circ \pm 4^\circ$ in very good
agreement with the results of, e.g., \citet{Marel01b}, who used red giants (Table~\ref{inclination_table}). \citet{Pejcha09} infer from RR~Lyrae stars that the eastern part is closer than the western part, but do not quantify the inclination angle. 

Along the short north-south axis covered by OGLE both the RR~Lyrae and Cepheid distributions show very similar behavior. For the Cepheids no inclination can be observed, while there may be a weak trend for the RR~Lyrae stars. Using a linear fit of the RR~Lyrae distribution we find the southern part of the OGLE~III region to be $0.9 \pm
0.2$~kpc closer than the northern part.

\subsection{Position angle}
\label{PA}

For the evaluation of the position angle of the LMC we count the number of all RR~Lyrae stars and separately of all Cepheids in boxes of $0.3^\circ \times 0.3^\circ$ in right ascension and declination. We determine the boxes with the highest number density in annuli of $0.3^\circ$ and fit these boxes with a first-order polynomial. 

The position angle varies somewhat depending on the location where it is measured. We find $\theta_{\mathrm{RRL}} = 102^\circ \pm 21^\circ$ and $\theta_{\mathrm{Cep}} = 113^\circ \pm 28^\circ$ if we take into account only the most populated areas in the innermost $3^\circ$ from the visual center of the LMC, as found also by \citet{Vaucouleurs72}. For the fields in a ring within a distance of $3^\circ < \mathrm{center} < 7^\circ$ we find $\theta_{\mathrm{RRL}} = 122^\circ \pm 32^\circ$ and $\theta_{\mathrm{Cep}} = 116^\circ \pm 25^\circ$. Thus for both tracer populations the position angle is found to increase at larger radii, indicative of mild twisting. The mean position angle over the entire OGLE~III field of the LMC is found to be $\theta_{\mathrm{RRL}} = 114^\circ \pm 13^\circ$ and $\theta_{\mathrm{Cep}} = 116^\circ \pm 18^\circ$. The number of Cepheids is much smaller, hence the statistics are worse and the uncertainty larger than for the RR~Lyrae stars. 

Our results are in good agreement with the values found for the LMC RR~Lyrae stars. For instance, \citet{Pejcha09} obtained $\theta_{\mathrm{RRL}} = 112.4^\circ$. Our results also agree well with what was found for other types of stars such as red giants or red clump stars (see Table~\ref{inclination_table}). The change of the position angle as a function of the position of the LMC was also pointed out by \citet{Marel01a} using red giants.

\subsection{Depth}

\begin{figure}
\centering 
 \includegraphics[width=0.47\textwidth]{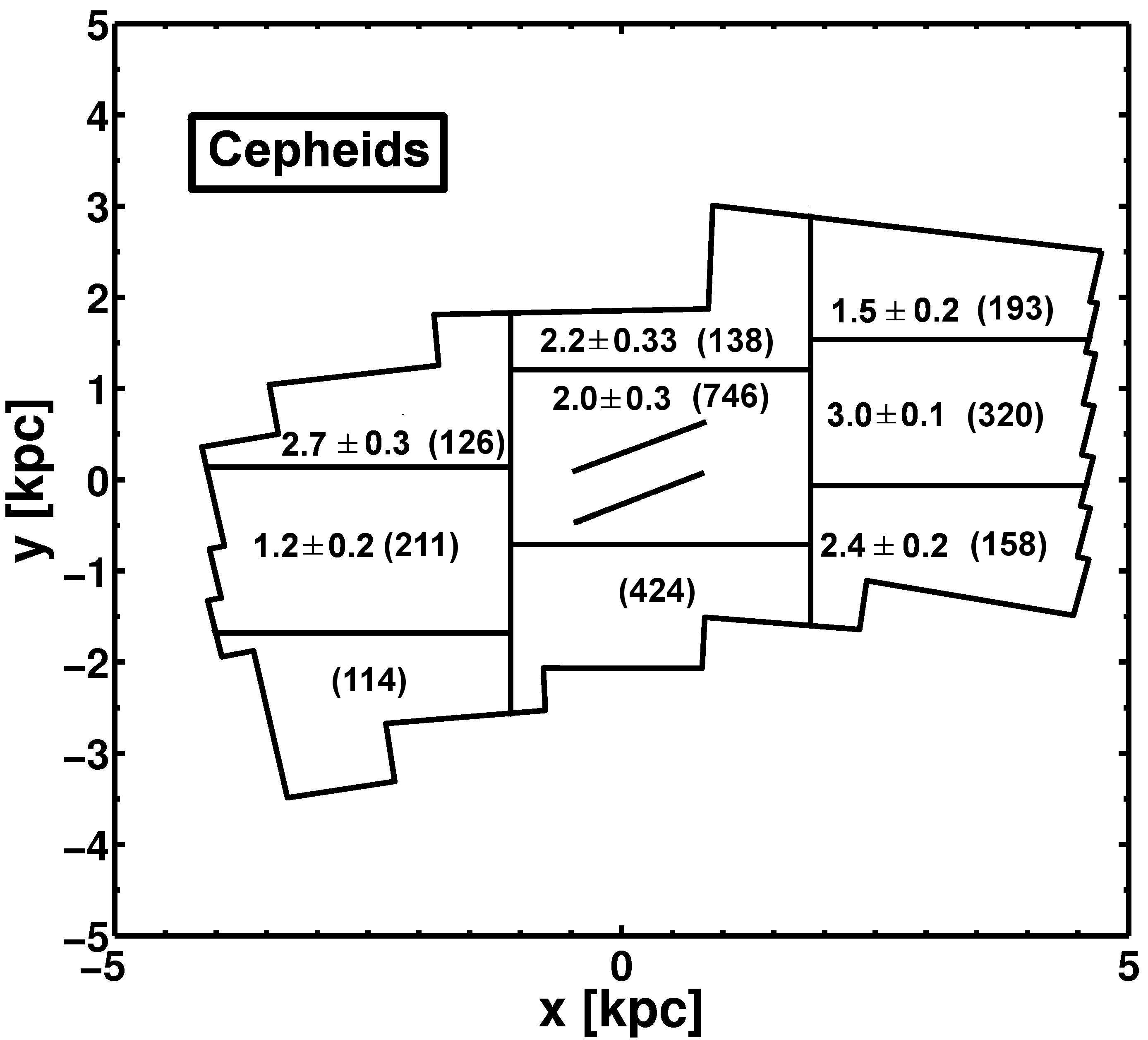}
\vspace{\floatsep} 
 \includegraphics[width=0.47\textwidth]{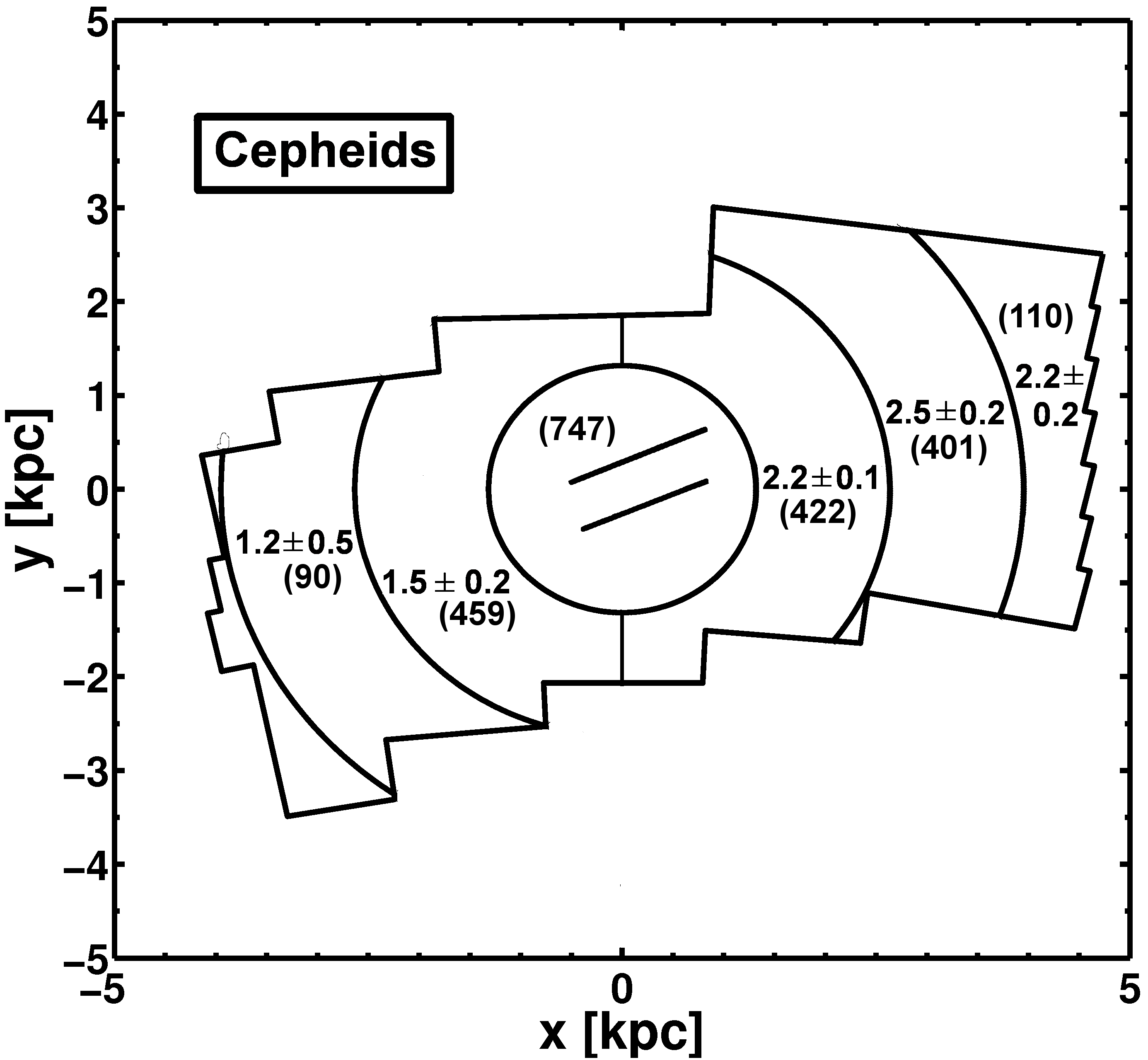}
\caption{Depth measurements for Cepheids in the LMC in fields of different projected sizes and locations, corrected for the inclination angle and position angle of the LMC. In the upper panel the OGLE~III field is divided into nine similarly sized roughly rectangular fields (truncated by the outer boundaries of the OGLE~III field). In the lower panel the depth is measured in four rings around the center of the bar. The annuli around the inner circle are furthermore divided into subfields containing stars with only positive or only negative $x$-coordinate values about the adopted center. The red numbers denote the deprojected depth of the Cepheids of the particular field in kpc. The depth has been computed by subtracting the quadrature of the uncertainty of the distance estimates from the quadrature of the raw depth. The depth of the Cepheids for some fields is smaller than the uncertainty. Therefore no depth estimate is shown for these fields. Underneath the depth values and their uncertainties the number of Cepheids used for each measurement is listed in brackets.} 
 \label{depth_LMC} 
\end{figure}

Before we proceed to determine the depth and scale height of the LMC we deproject the spatial positions of the stars by the inclination and position angle determined previously. The deprojection is done with respect to the center of the RR~Lyrae star distribution at $\alpha = 5^h26^m$ and $\delta = -69^\circ75'$. Using the approach of \citet{Weinberg01} we determine the $x$, $y$, and $z$ coordinates of every star and use two different sets of fields to measure the depth and scale height of the LMC (Figure \ref{depth_LMC}). 

The density distribution of the RR~Lyrae stars changes across the OGLE~III field (see Figure~\ref{RRL_Cep_RADEC_colourext}) and we want to approximate the isodensity contours. To do so the OGLE~III field is divided into nine rectangular fields of similar size (unless truncated by the outer boundaries of the OGLE~III field), as well as into four rings with respect to the origin of the new coordinate system. These fields are shown in the upper and lower panel of Figure~\ref{depth_LMC}). For each field we calculate a cumulative distribution function of all stars present, as shown in Figure~\ref{cum_dist}. The depth is calculated by measuring the minimal and maximal distances of the central 68\% of the distribution. This is accomplished by computing the distance where 16\% and 84\% of the whole sample of stars have shorter distances than a particular star. Using this approach we become independent of the possible asymmetric geometry of the LMC, as observed in the bar region. To estimate the uncertainty of the calculations of the depth, we vary the central 68\% of our cumulative distribution by 5\% in positive and negative $z$ direction. The mean of the depth differences is assumed to be the uncertainty of each field.

These depth estimates are of the same order as the distance uncertainties of the individual stars. Assuming a flat disk with zero depth of the stars together with the distance uncertainties could lead to a similar distribution of stars as observed for the LMC. Therefore we have to subtract the quadrature of two times the mean distance uncertainty, denoted as $(2\sigma_{star})^2$, of each stellar population from the quadrature of the raw depth estimate that we observe. We choose $2\sigma$ here, because our definition of the depth is equivalent to two standard deviations from the mean. For the Cepheids a mean distance uncertainty of $1.8$~kpc is obtained, while the RR~Lyrae star distances have a mean uncertainty of $3.2$~kpc since here the uncertainties of the photometric metallicities enter as well. The ``true'' depth of the distribution is calculated with
\begin{equation}
depth_{true} = \sqrt{depth_{raw}^2 - (2\sigma_{star})^2}
\label{real_depth}
\end{equation}

The number of RR~Lyrae stars per field as well as the raw depth of the old population of the LMC varies within the OGLE field, as shown in Figure~\ref{RRL_Cep_RADEC_colourext}. In Figure~\ref{depth_LMC}, we subdivide the OGLE~III field into rectangular and semi-annular fields. For fields with rectangular inner boundaries we find raw depth values ranging from from 4.4~kpc to 5.4~kpc with a mean value of $4.9 \pm 0.2$~kpc. The annular fields are subdivided by considering only stars with either positive or only negative values of their $x$ coordinate. With this approach raw depth values between 4.5~kpc to 5.5~kpc are found for these fields. The mean value is $5.0 \pm 0.2$~kpc. However, all these depth estimates are not significant, because the distance uncertainty $\sigma_{star}$ is larger than the depth and Equation~\ref{real_depth} results in a negative value. Therefore, we cannot give a significant result for the depth of the RR~Lyrae stars.

Although the density of the Cepheids is much lower than that of the RR~Lyrae stars, the Cepheid distance errors are also much smaller, and we evaluate their depth in the same manner and for the same fields as for the RR~Lyrae stars. The raw depth values measured in the field in Figure~\ref{depth_LMC} range from 3.3~kpc to 4.7~kpc. Correcting for the uncertainty of the distance estimates depth values up to 3.0~kpc are found as shown in Figure~\ref{depth_LMC} (upper panel). The mean depth is $1.7 \pm 0.2$~kpc. For the annular fields in the lower panel of Figure~\ref{depth_LMC}, raw depth values between 3.6~kpc to 4.4~kpc are measured. The corrected depth values are therefore up to 2.5~kpc (as listed in this panel). The mean depth for the annular fields is $1.7 \pm 0.2$~kpc. For some fields the raw depth is smaller than the uncertainty, and therefore some of the fields do not have an estimate for the depth. Overall we observe slightly higher depth values for Cepheids in the western semi-annular fields of the OGLE~III area. 

For comparison, we note that a mean raw depth of $9.4 \pm 1.3$~kpc is found when applying the area-averaged RC reddening to RR~Lyrae stars, and a mean raw depth of $10.4 \pm 1.0$~kpc then results for Cepheids. These values have to be corrected for the distance uncertainties of $3.9$~kpc for the RR~Lyrae stars and $2.2$~kpc for the Cepheids. The mean depth values are $5.2$~kpc for the RR~Lyrae stars and $9.4$~kpc for the Cepheids. Both values are significantly higher than the depth obtained with individual de-reddening. The large difference between the area-averaged and individual reddening-corrected depth for the Cepheids may imply that Cepheids are more strongly affected by differential reddening than RR Lyrae stars. This could be due to their being associated with regions of more recent star formation or due to effects of circumstellar dust around these extended supergiants \citep{Barmby11}.

\subsection{Scale Height}
\label{scale_height}

The scale height of a disk is defined as the distance where the density of stars has decreased by a factor of $1/e$. Using the same approach as for calculation of the depth two times the scale height would correspond to a central width of 63.2\% in the cumulative distribution of each field. Therefore we can calculate the scale height by multiplying the depth by a factor of 

\begin{equation}
 \frac{\textrm{scale~height}}{\textrm{depth}} = 
\frac{1 - 2(\frac{1}{2e})}{2 \times 0.68} = 0.4648.
\end{equation}

Using this formula we estimate mean scale heights of $0.8 \pm 0.2$~kpc for the Cepheids. 

In \citet{Subramanian09} the depth of the RC stars is determined for the OGLE~II fields, which cover only the most central parts of the LMC. This study finds a mean depth of $4.0 \pm 1.4$~kpc for the bar region. In \citet{Subramaniam09a} the scale height of the RR~Lyrae stars for the OGLE~III field is investigated, but due to large uncertainties no value is determined, as in our study. 

The scale height of the Cepheids is significantly higher than the 180 pc determined for the H~{\sc i} scale height of the LMC by \citet{Kim99}. \citeauthor{Kim99}'s value is based on the H~{\sc i} velocity dispersion as measured in ``relatively quiescent regions''. If we assume instead an H~{\sc i} velocity dispersion of 15.8~km~s$^{-1}$ (\citealt{Kim98b}; see also \citealt{Meatheringham88}), the resulting scale height of the neutral hydrogen would be of the order of 800~pc when using Equation~1 of \citet{Kim99} in very good agreement with our result.

As pointed out by \citet{Kim07} and other authors the H~{\sc i} distribution in the LMC has a complex, seemingly perturbed structure consisting of clumps and holes. (Similar behavior has been found in other irregular galaxies.) The distribution of the Cepheids and other, more recently formed stars mimics such an irregular distribution, where we find clumpy density enhancements of young stars at different distances, or regions with little recent activity. The loci of the Cepheids still trace the approximate loci of the associations and clusters in which they formed originally. Perhaps this irregularly distributed star formation is contributing to the higher inferred scale height of the Cepheids.

\citet{Olsen11} suggest that the interactions between the Magellanic Clouds lead to the infall of gas (and stars) in the LMC. This infall may enhance the scale height and trigger further star formation. Interestingly, it was found that the red supergiants in the LMC (which may be in a similar age range as the Cepheids) rotate with a higher velocity than the H~{\sc i} \citep{Olsen07}. Also, \citet{Olsen07} identify stars that appear to be associated with the H~{\sc i} tidal arms in the LMC rather than to participate in ordered disk rotation. \citet{Olsen11} discuss that there appears to be a (small) population of (primarily AGB) stars that is counterrotating with respect to the disk of the LMC; stars that may belong to a plane inclined by $54\degr$ w.r.t.\ the disk of the LMC. As pointed out earlier the dynamical center of the H~{\sc i} is offset from the photometrically inferred center of the stellar distribution \citep[e.g.,][]{Cole05}). While these recent findings may not necessitate as high a scale height as we found for the Cepheids, they do suggest a much more complex structure of the LMC than previously anticipated.

The analysis of \citet{Casetti12} of red giants and supergiants in the LMC support \citet{Olsen11} suggestions that these stars may have been captured from the SMC. \citeauthor{Casetti12}'s proper motion study suggests that these stars lie in an inclined prograde disk w.r.t. the LMC's main stellar distribution. Furthermore, these authors propose that the most recent LMC-SMC interaction took place some 200~Myr ago. All this may contribute to the overall high Cepheid scale height and the even slightly larger scale heights found in our data in the western part of the LMC.

%

\section{Summary and Conclusions}
\label{Conclusions}

\begin{table*}
\begin{center}
\caption{Distances and structural parameters for the LMC derived in this paper.} 
 \label{Summary_table} 
\centering 
\begin{tabular}{r c c} 
\tableline\tableline 
 & RR~Lyrae & Cepheids \\ 
\tableline 
Distance [kpc] (area-averaged de-reddening) & $52.7 \pm 3.9$ & $58.8
\pm 2.2$ \\ 
Distance [kpc] (individual de-reddening) & $53.1 \pm 3.2$ & $53.9 \pm 1.8$ \\ 
Inclination angle [degrees] & $32 \pm 4$ & $32 \pm 4$ \\ 
Position angle [degrees] & $114 \pm 13$ & $116 \pm 18$ \\ 
Position angle [degrees] (innermost $3^{\circ}$) & $102 \pm 21$ & $113
\pm 28$ \\ 
Position angle [degrees] ($3^{\circ}$ to $7^{\circ}$) & $122 \pm 32$ &
$116 \pm 25$ \\ 
Mean depth [kpc] & \nodata & $1.7 \pm 0.2$ \\ 
Scale height [kpc] & \nodata & $0.8 \pm 0.2$ \\
\tableline  
\end{tabular}
\end{center}
\end{table*}

We calculate distances to individual RR~Lyrae stars observed by the OGLE~III survey using the metallicity estimates based on the Fourier decomposition of their light curves \citep{Haschke12_MDF}. Additionally we use the OGLE~III Cepheids to determine distances to the young population of the LMC. 

Extinction corrections are applied using two different techniques presented in \citet{Haschke11_reddening}. In the first approach we calculate the mean reddening of a field based on the mean color of the RC stars contained in that field. In the second approach we calculate individual reddening values for each RR~Lyrae star and Cepheid by comparing the absolute magnitudes of the stars with the observed magnitudes. With this second technique we are able to correct the actual reddening at the position of the star. We create a self-consistent three-dimensional map of the LMC for RR~Lyrae and Cepheid stars using both de-reddening techniques.

When using the individual reddening corrections we obtain a median distance of $D_{RRL/median} = 53.1 \pm 3.2$~kpc for the RR~Lyrae stars and of $D_{\mathrm{Cep/median}} = 53.9 \pm 1.8$~kpc for the Cepheids, i.e., the distances agree very well within the uncertainties. If we use the average reddening values from the RC stars instead, the median locus of the RR~Lyrae stars is 6~kpc closer than that of the Cepheids (see Table~\ref{Summary_table}). 

Our investigation thus suggests a possible solution of the long and short distance scale problem for RR~Lyrae and Cepheid distances in the LMC. The discrepancy of distance moduli from RR~Lyrae stars and Cepheids has been known for many decades \citep[e.g.,][]{Tammann77}. The distances of Population~II stars are on average shorter than the distances resulting from Population~I stars \citep{Clementini03}, and this discrepancy is often larger than the uncertainties of the individual estimates. With our result of $\Delta(m-M) = 0.03$~mag for the individually reddening-corrected distances this difference is much reduced as compared to many earlier studies and smaller than the uncertainties of the distance estimates. 

We suggest that the long and short distance scale problem may be related to the extinction correction. In \citetalias{Haschke11_reddening} we have shown that the reddening estimates of different authors using different techniques can be significantly different. This may in part be caused by the use of averaged reddening estimates which necessarily cannot account for differential reddening on small scales. Moreover, extinction varies as a function of stellar population \citep[e.g.,][]{Grebel95, Zaritsky04}. Cepheids may experience considerable mass loss, which leads to a light-absorbing gas reservoir in the vicinity of the star. This effect may dim the star light and, if unrecognized, lead to a larger distance. Individual de-reddening therefore leads to more reliable results (as also evidenced by the more pronounced clustering of Cepheids described in Section~\ref{reddening_test}). 

The distribution of stars in the LMC can be described as an inclined disk \citep[e.g.,][]{Marel01a}. The gas, traced by H~{\sc i}, is rotating in a disk as well \citep[e.g.,][]{Kim98a}, while the dynamical centers of the stellar and the gaseous distribution are separated by $1.2^{\circ}$ \citep{Marel01a}. \citet{Borissova04, Borissova06} measured the velocity dispersion of RR~Lyrae stars along and below the LMC bar. Their rather high velocity dispersion of $\sim 50$~km~s$^{-1}$ suggests the presence of a kinematically hot halo, yet to be confirmed by measurements of RR~Lyrae stars farther away from the LMC center. Red giants show a much smaller velocity dispersion of 24.7~km~s$^{-1}$ and are confined to the disk \citep{Cole05}.

Using the individually reddening-corrected distances we obtain an inclination angle of $i_{\mathrm{RRL,Cep}} = 32^\circ \pm 4^\circ$ for RR~Lyrae stars and Cepheids, with the eastern part of the LMC closer to us than the western part. For the position angle of the RR~Lyrae stars we find a dependence on galactocentric distance. Overall we get $\theta_{\mathrm{RRL}} = 114^\circ \pm 13^\circ$ for the RR~Lyrae stars and $\theta_{\mathrm{Cep}} = 116^\circ \pm 18^\circ$ for the Cepheids, respectively. For the innermost 3~kpc from the optical center of the LMC we obtain $\theta_{\mathrm{RRL}} = 102^\circ$ and for the outer regions in a distance from 3~kpc to 7~kpc from the optical center $\theta_{\mathrm{RRL}} = 122^\circ$. Overall our results of the structural parameters of the LMC are in very good agreement with the literature (Table~\ref{inclination_table} and Table~\ref{Summary_table}). Our results thus confirm earlier findings of an inclined LMC disk. Moreover, the change of position angle -- seen both in tracers of the old and of the young population -- suggest a twisted or warped disk, which may be the result of interactions between the Magellanic Clouds and the Milky Way. Indications for a warped disk were also found in other studies, e.g., by \citet{Marel01b} from RGB stars. We note that the OGLE data, while covering the main body of the LMC, do not cover its full extent. Hence, based on our data alone, we can not investigate the behavior in the LMC's outskirts. 

The depth of the LMC RR~Lyrae stars cannot be determined with sufficient significance, because the distance uncertainties of the RR~Lyrae stars are too large. For the depth of the Cepheids values up to 3.0~kpc are found and a mean depth of $1.7 \pm 0.2$~kpc is obtained. The mean scale height of the Cepheids is $0.8 \pm 0.2$~kpc. 

The RR~Lyrae stars show an extended smooth distribution with a pronounced central concentration in the LMC, while the clumpy distribution of the Cepheids still traces their irregularly distributed formation sites. When using intermediate-age and older stars such as AGB and RG stars, the LMC shows a smooth disk and a pronounced bar \citep{Marel01a}. A similar picture can be seen for the RR~Lyrae stars in our data. 

The distance and extent of the off-centered bar region of the LMC differs from that of the main disk. The density of stars located more than $1\sigma$ in front of the median distance of the RR~Lyrae stars is quite low compared with the main locus of the disk of the LMC, but in the bar region the density of stars is about $50\%$ higher than in the other regions covered by OGLE~III. We find that for the RR~Lyrae stars the bar protrudes out of the disk by $\sim 5$~kpc. This is of the same order as the depth of the LMC RR~Lyrae stars. For the Cepheids we do not find the bar to be such an outstanding feature. 

A number of different explanations for the unusual structure of the bar have been discussed in the literature. For instance, the off-center location compared to the optical and kinematical center \citep{Marel01b} and the extent outwards of the disk \citep{Nikolaev04} led to the suggestions that the LMC might have collided with a small galaxy \citep{Zhao00} or with a dark halo with a few percent of the mass of the LMC \citep{Bekki09b}. As discussed by \citet{Casetti12}, simulations of the interaction of a barred disk galaxy with a smaller companion lead to a temporarily off-centred bar in the case of a major-axis encounter \citep[see simulations by][]{Berentzen03}. \citeauthor{Berentzen03} also describe the resulting thickness of the disk and density enhancements in ring-like structures and at the end of the bar in qualitative support of our enhanced scale heights. \citet{Zaritsky04b} suggested that the bar might be a bulge, which does not seem to be consistent with our findings. Interestingly \citet{Subramaniam09b} did not find evidence for a bar located in front of the LMC using the red clump data of OGLE~III. Based on our current study, we conclude that the bar of the LMC is a well-defined structure partially in front of the LMC disk. Dependent on the stellar tracers the bar feature has different characteristics but is well visible in both old and young populations.

%
%

\acknowledgments

We are thankful to the OGLE collaboration for making their data publicly available. R.\ Haschke is obliged to S.\ Jin and S.\ Pasetto for giving helpful comments on the mathematical problems arising during this project. The comments to improve the manuscript and the proof reading by K.\ Glatt and K.\ Jordi are very much appreciated. This work was partially supported by Sonderforschungsbereich SFB~881 ``The Milky Way System'' (subproject A2) of the German Research Foundation (DFG). Finally, we thank an anonymous referee for helpful comments that improved this paper.

\bibliography{Bibliography.bib}
\bibliographystyle{apj}

\end{document}